\documentclass{article}

\PassOptionsToPackage{numbers,compress}{natbib}
\usepackage[dblblindworkshop,final]{neurips_2025}

\workshoptitle{Data on the Brain \& Mind}

\usepackage[utf8]{inputenc} 
\usepackage[T1]{fontenc}    
\usepackage{hyperref}       
\usepackage{url}            
\usepackage{booktabs}       
\usepackage{amsfonts}       
\usepackage{nicefrac}       
\usepackage{microtype}      
\usepackage{xcolor}         

\usepackage{booktabs}
\usepackage{longtable}
\usepackage{amsmath}
\usepackage{graphicx}
\usepackage{subcaption}
\usepackage{multirow}
\usepackage[export]{adjustbox}
\usepackage[linesnumbered, ruled, vlined]{algorithm2e}
\usepackage{pifont} 
\usepackage{bbm}

\usepackage{tikz}

\usetikzlibrary{arrows.meta, positioning}

\title{SpellerSSL: Self-Supervised Learning with P300 Aggregation for Speller BCIs}

\author{%
  Jiazhen Hong\thanks{Work done during internship at Emotiv Research.} \\
  Emotiv Research \\
  Melbourne, Australia \\
  \texttt{jiazhen@emotiv.com} \\
  \And
  Geoffrey Mackellar \\
  Emotiv Research \\
  Sydney, Australia \\
  \texttt{geoff@emotiv.com} \\
  \And
  Soheila Ghane \\
  Emotiv Research \\
  Melbourne, Australia \\
  \texttt{soheila@emotiv.com} \\
}

\begin{document}

\maketitle
\begin{abstract}
Electroencephalogram (EEG)-based P300 speller brain–computer interfaces (BCIs) face three main challenges: low signal-to-noise ratio (SNR), poor generalization, and time-consuming calibration. We propose \textit{SpellerSSL}, a framework that combines self-supervised learning (SSL) with P300 aggregation to address these issues. First, we introduce an aggregation strategy to enhance SNR. 
Second, to achieve generalization in training, we employ a customized 1D U-Net backbone and pretrain the model on both cross-domain and in-domain EEG data. The pretrained model is subsequently fine-tuned with a lightweight \textit{ERP-Head} classifier for P300 detection, which adapts the learned representations to subject-specific data. 
Our evaluations on calibration time 
demonstrate that combining the aggregation strategy with SSL significantly reduces the calibration burden per subject and improves robustness across subjects.
Experimental results show that SSL learns effective EEG representations in both in-domain and cross-domain, with in-domain achieving a state-of-the-art character recognition rate of 94\% with only 7 repetitions and the highest information transfer rate (ITR) of 21.86~bits/min on the public II-B dataset. Moreover, in-domain SSL with P300 aggregation reduces the required calibration size by 60\% while maintaining a comparable character recognition rate. 
To the best of our knowledge, this is the first study to apply SSL to P300 spellers, highlighting its potential to improve both efficiency and generalization in speller BCIs and paving the way toward an EEG foundation model for P300 speller BCIs. 
The code is available at: https://github.com/Emotiv/SpellerSSL-NeurIPS2025
\end{abstract}


\section{Introduction}
Electroencephalogram (EEG)-based Brain–Computer Interfaces (BCIs), due to their non-invasive nature, low cost, and ease of use, have been widely adopted as assistive technologies for individuals with motor disabilities \cite{lotte2018review}. Among various paradigms, the P300 is an event-related potential (ERP) component that typically occurs around 300~ms after a target stimulus in an oddball paradigm \cite{bacsar1984new}. It has been widely applied in speller BCI, known as the P300 Speller (visualized in Figure~\ref{fig:1}(a)).  

Traditional P300 speller BCIs face three main challenges:  
(1) \textbf{Low signal-to-noise ratio (SNR):} The low SNR of EEG signals makes reliable single-trial P300 detection difficult \cite{hong2024p3t}. Current systems often require multiple repetitions of each stimulus to achieve an accurate character recognition rate (CRR), which slows down spelling speed and reduces the information transfer rate (ITR) \cite{wang2023st}.  
(2) \textbf{Poor generalization:} Existing methods used in P300 spellers include conventional machine learning algorithms such as stepwise linear discriminant analysis (SWLDA) \cite{speier2018improving} and support vector machines (SVMs) \cite{akram2013novel}, as well as deep learning architectures such as convolutional neural networks (CNNs) \cite{cecotti2010convolutional}, CapsNet with attention \cite{ma2021capsule, wang2023st}, stacked Transformers \cite{hong2024p3t}, and image-based representations such as TimeSformer \cite{hong2025topoeeg}. However, these approaches are task- or subject-specific and fail to provide generalizable representations of the P300 response.  
(3) \textbf{Time-consuming calibration:} Large inter-individual differences require a time-consuming and subject-specific calibration session before the speller can be used in real-time and online \cite{hong2024chatbci, chandravadia2025comparing}.    

In brief, the three main questions are: (1) how to obtain higher signal quality, (2) how to build more generalizable models, and (3) how to reduce calibration requirements.  

Recent advances in self-supervised learning (SSL) and foundation models offer promising directions for addressing these challenges. In this work, we propose \textit{SpellerSSL}, a P300 speller framework that integrates SSL with P300 aggregation strategies. Our contributions are summarized as follows:

\begin{itemize}
    \item We introduce a P300 aggregation strategy that enhances signal quality in the training set, enabling the model to learn more reliable single-trial responses and achieve higher character recognition rates (CRRs) with fewer repetitions.  
    \item We customize a 1D U-Net~\cite{ronneberger2015unet} as the backbone SSL model and show that an SSL model pretrained on both cross-domain and in-domain EEG data can enhance the generalization of learned representations. 
    We also design a lightweight \textit{ERP-Head} classifier for downstream detection of P300, which allows subject-specific adaptation.  
    \item We evaluate calibration efficiency under reduced training set sizes and demonstrate that combining the aggregation strategy with a pretrained model enhances robustness, reducing calibration time without compromising accuracy in P300 speller BCIs.  
\end{itemize}

To the best of our knowledge, this is the first work to explore SSL in P300 speller BCIs. We demonstrate that the proposed SpellerSSL improves both efficiency and generalization, paving the way for the future deployment of foundation models in practical P300 speller systems.

\section{Method}
\begin{figure}
  \centering
  \includegraphics[width=0.85\linewidth]{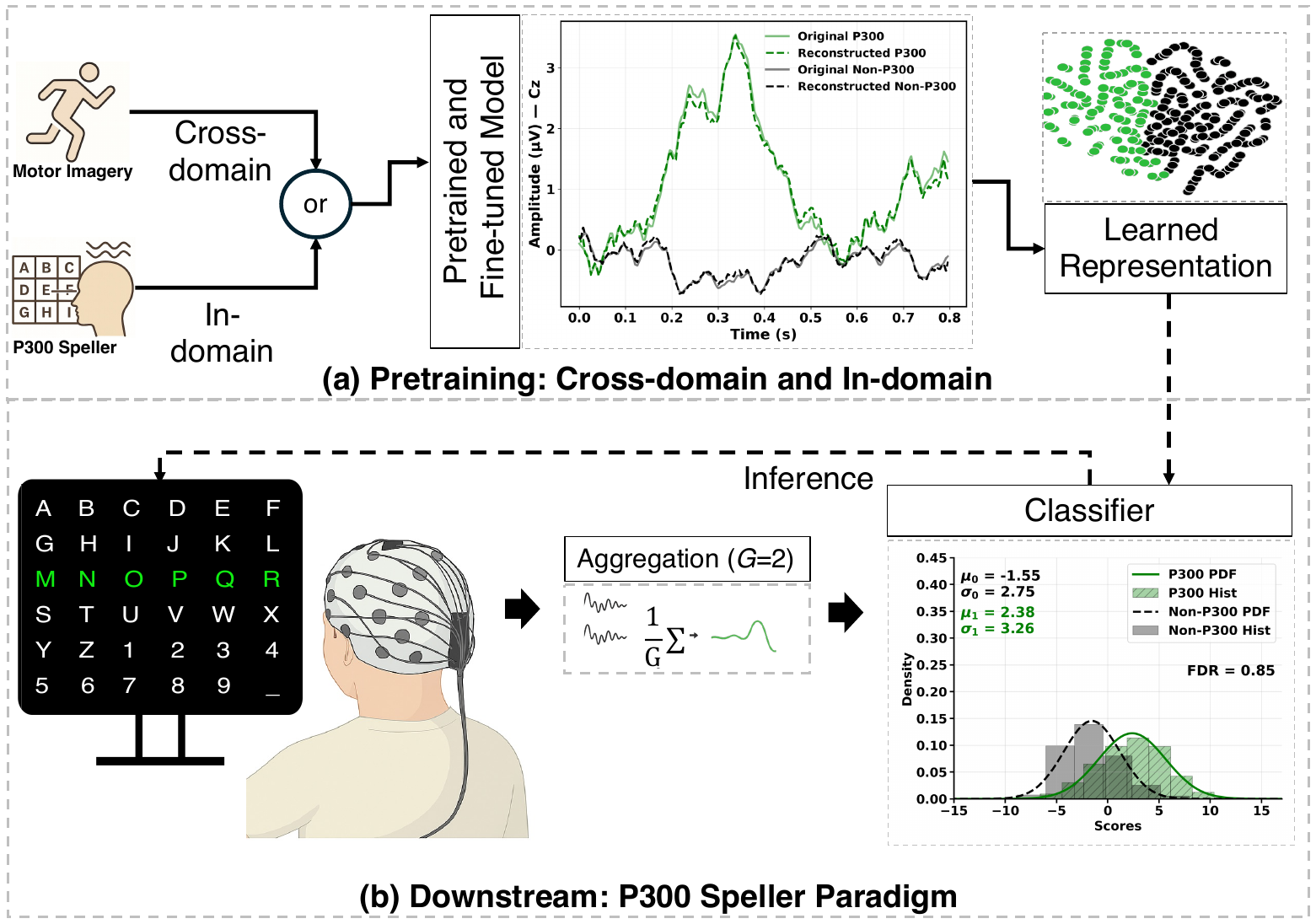}
  \caption{Overview of the proposed SpellerSSL. 
  }
  \label{fig:1}
\end{figure}

Figure~\ref{fig:1} provides an overview of SpellerSSL. 
In (a), self-supervised pretraining is performed with a full 1D U-Net backbone via a reconstruction task on either cross-domain or in-domain EEG (Section~\ref{sec:M1}). 
In (b), downstream P300 detection is performed, where the learned representations at the U-Net bottleneck layer are fine-tuned with an ERP-Head classifier (Section~\ref{sec:M2}). 
A P300 aggregation step, averaging $G$ consecutive repetitions, is applied during fine-tuning to enhance the quality of single-trial signals (Section~\ref{sec:M3}).

\subsection{Pretraining} \label{sec:M1} 
We customized 1D U-Net architecture (Appendix~\ref{app:unet}) such that input sequences are zero-padded in time so that output length is a multiple of 16, 
matching the encoder’s four successive downsamplings (each with stride~2).
The model is trained for 200 epochs with a batch size of 64, using AdamW (initial learning rate \(2.5\times10^{-4}\), weight decay \(1\times10^{-2}\)) and a OneCycle schedule~\cite{smith2019super} (maximum learning rate \(5\times10^{-4}\), final learning rate \(5\times10^{-6}\), 10\% warmup, cosine decay). We optimize the model to reconstruct randomly masked EEG sequences (50\% time masking, 0\% channel masking).

Let the input EEG trial be \(x\in\mathbb{R}^{C\times L}\). 
The masked input \(\tilde{x}\) is passed through the model \(f_\theta(\cdot)\) to obtain the reconstruction \(\hat{x}=f_\theta(\tilde{x})\). 
We adopt a time–frequency consistency objective:
\begin{equation}
\mathcal{L}_{\mathrm{SSL}} 
= \|x - \hat{x}\|_{1} \;+\; \lambda \,\big\|\mathrm{FFT}(x) - \mathrm{FFT}(\hat{x})\big\|_{1},
\label{eq:ssl}
\end{equation}
where $\mathrm{FFT}(\cdot)$ denotes the fast Fourier transform. The frequency-domain term encourages the encoder to capture spectral characteristics of EEG signals, complementing the time-domain reconstruction.

\subsection{Downstream}\label{sec:M2}
After training, we extract the encoder’s bottleneck feature map before upsampling,
\(
b \in \mathbb{R}^{1024 \times \tfrac{L}{16}},
\)
and discard the decoder. The ERP-Head \(\mathrm{Head}_\phi(\cdot)\) operates directly on the sequence \(b\) and applies the following transformations:
\begin{align}
\tilde{b} &= \mathrm{PointwiseConv}(b)\ \xrightarrow{}\ \mathrm{DepthwiseConv}\ \xrightarrow{}\ \mathrm{DilatedDepthwiseConv}\ \xrightarrow{}\ \mathrm{PointwiseConv}, \\
h &= \mathrm{GAP}(\tilde{b}) \in \mathbb{R}^{D}, \qquad 
z = W h + c \in \mathbb{R}^{2},
\end{align}
where PointwiseConv denotes a $1\times1$ projection and DepthwiseConv operates on temporal dimension.

We minimize the supervised cross-entropy loss
\begin{equation}
\mathcal{L}_{\mathrm{sup}} = \mathrm{CE}(z, y),
\end{equation}
where $\mathcal{L}_{\mathrm{sup}}$ denotes the supervised loss, and $\mathrm{CE}$ is the standard cross-entropy between predicted logits $z$ and ground-truth labels $y$. A fixed positive vs. negative class weight ratio of \(1:5\) is used to address class imbalance. 
When an SSL checkpoint is available, encoder-only weights up to the bottleneck are loaded, and the model is fine-tuned for 10 epochs on the training set. 
The detailed architecture of the ERP-Head is provided in Appendix~\ref{app:erphead}.

\subsection{P300 Aggregation}\label{sec:M3}
\begin{figure}
  \centering
  \includegraphics[width=\linewidth]{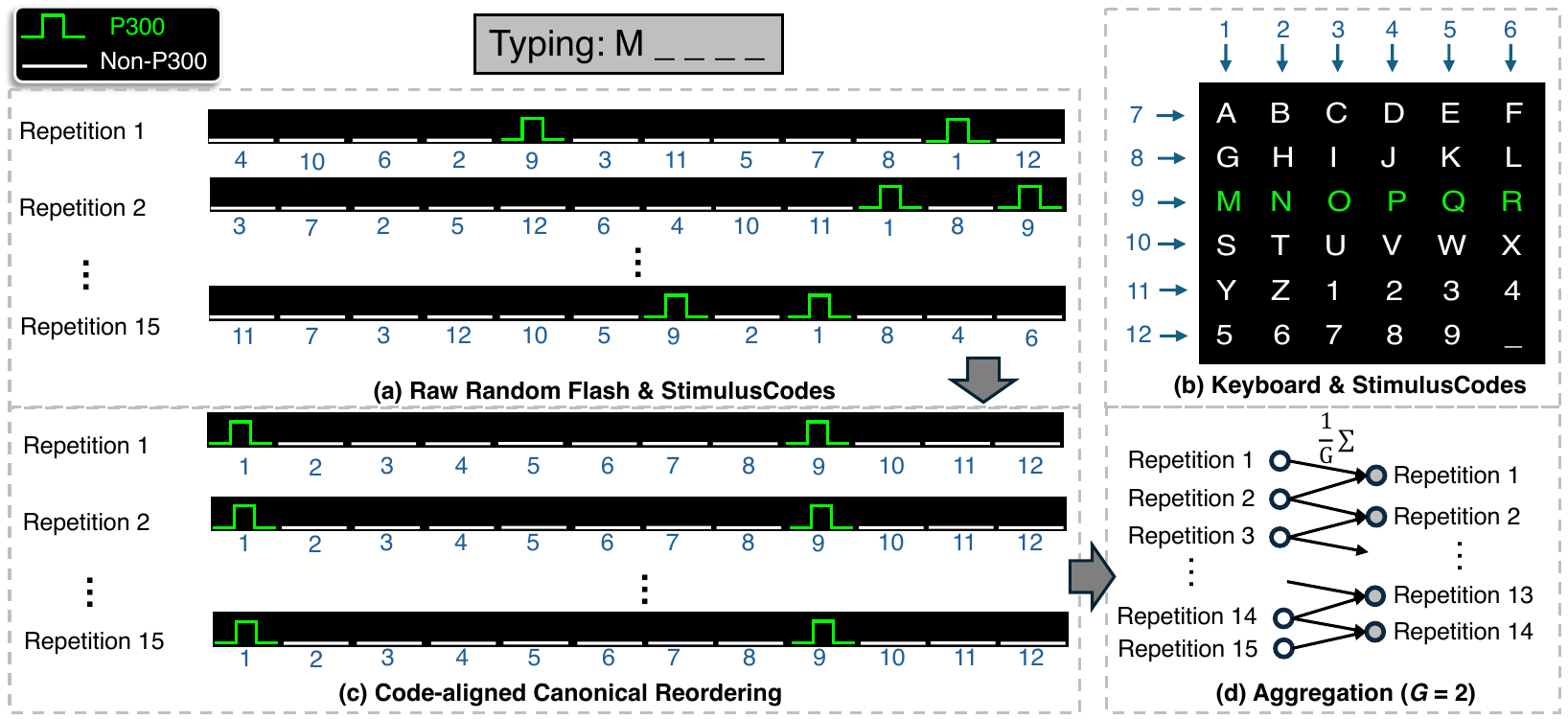}
  \caption{P300 aggregation for typing the character ``M''.
  }
  \label{fig:2}
\end{figure}

Figure~\ref{fig:2} summarizes the proposed aggregation pipeline for typing the character ``M''. 
Panel (a) shows the raw random flashes with their \textit{StimulusCodes}~\cite{krusienski2004bci} over $R{=}15$ repetitions; green segments denote P300 (row/column hits), while white segments denote Non-P300 responses.
Panel (b) illustrates the $6{\times}6$ speller with its code assignment; for ``M'', the target row/column are $(\text{row }3,\ \text{col }1)$, corresponding to codes $(9,1)$.
Panel (c) performs code-aligned canonical reordering within each repetition; the 12 trials are permuted to a fixed order $k{=}1..12$ by aligning to the 12 StimulusCodes.
Panel (d) applies a sliding-window aggregation with group size $G{=}2$ (with overlap), where every two consecutive repetitions are averaged to form aggregated trials.

\paragraph{Notation.}
Each character selection is presented over $R$ repetitions (here $R{=}15$), with 12 row/column flashes per repetition. 
Let $x^{(r)}_{k}\in\mathbb{R}^{C\times L}$ denote the EEG trial in repetition $r\!\in\!\{1,\dots,R\}$ corresponding to the flash with canonical code index $k\!\in\!\{1,\dots,12\}$ (to be aligned in Step 1). 
Let $s^{(r)}_j\!\in\!\{1,\dots,12\}$ denote the StimulusCode of the $j$-th flash in temporal order within repetition $r$. 
For the typed character, define the P300 code set 
\begin{equation}
\mathcal{K}_{\mathrm{P300}} = \{k_{\mathrm{row}},\, k_{\mathrm{col}}\},
\quad 
k_{\mathrm{row}} \in \{7,\ldots,12\},\;
k_{\mathrm{col}} \in \{1,\ldots,6\}.
\end{equation}
which contains the row and column codes of that character (e.g., ``M'': $\mathcal{K}_{\mathrm{P300}}=\{9,1\}$). 
The per-code label is defined as
\begin{equation}
y[k] \;=\; 
\begin{cases}
1, & k \in \mathcal{K}_{\mathrm{P300}} \quad(\text{P300})\\[2pt]
0, & \text{otherwise} \quad(\text{Non-P300})
\end{cases}
\quad\text{for } k=1,\dots,12,
\end{equation}
so that each repetition contains exactly two P300 and ten Non-P300 flashes.

\paragraph{Step 1: Code-aligned canonical reordering.}
Within each repetition, we permute the 12 trials to a canonical order $k{=}1..12$ using the StimulusCode. 
Let $\pi_r$ map the canonical code index $k$ to its temporal position in repetition $r$, i.e.
\begin{equation}
s^{(r)}_{\pi_r(k)} \;=\; k',\qquad k'=1,\dots,12.
\end{equation}
We then form the ordered tensor
\begin{equation}
X^{(r)}_{\mathrm{ord}}[k'] \;=\; x^{(r)}_{\pi_r(k)} \;\in\; \mathbb{R}^{C\times L},
\qquad
X^{(r)}_{\mathrm{ord}} \;\in\; \mathbb{R}^{12\times C\times L},
\end{equation}
where the subscript ``ord'' denotes canonical ordering by StimulusCodes.

\paragraph{Step 2: Sliding-window aggregation (codes retained).}
We aggregate $G$ consecutive repetitions with a sliding window (overlap allowed), yielding windows indexed by $r=1,\dots,R-G+1$. 
For each window and canonical code $k$, we average trials with the same code $k$ across the $G$ repetitions:
\begin{equation}
\bar{X}^{(r,G)}_{\mathrm{ord}}[k']
\;=\;
\frac{1}{G}\sum_{t=r}^{r+G-1} X^{(t)}_{\mathrm{ord}}[k'],
\qquad
k'=1,\dots,12,
\end{equation}
so the aggregated window still contains 12 trials. 
The labels are preserved per code,
\begin{equation}
y^{(r,G)}[k'] \;\equiv\; y[k'] \;=\; 
\begin{cases}
1, & k' \in \mathcal{K}_{\mathrm{P300}} \quad(\text{P300})\\[2pt]
0, & \text{otherwise} \quad(\text{Non-P300})
\end{cases}
\end{equation}
which preserves the original speller paradigm and maintains the two-P300 vs. ten-Non-P300 pattern in every window. 
Aggregation is applied only to calibration (training) data and is not performed during real-time or online operation.

\section{Experiment}

\subsection{Dataset \& Pre-processing}
We use two publicly available datasets: PhysionetMI~\cite{goldberger2000eegmmidb, schalk2004bci2000} (motor imagery) and BCI Competition III-II~\cite{krusienski2004bci} (P300 speller).  
PhysionetMI, accessed via the MOABB pipeline~\cite{moabb2018}, serves as the cross-domain pretraining set.  
BCI Competition III-II is band-pass filtered at $0.1$–$60$ Hz and consists of two subjects, generally split into two datasets, II-A and II-B~\cite{wang2023st}.  
We use II-A for in-domain pretraining and II-B for downstream evaluation.

All datasets follow the 10–20 system with 64 channels in the same order. PhysionetMI provides 3-second trials at 160 Hz, while II-A/II-B provide 1-second trials at 240 Hz. To standardize, we resample all signals to 240 Hz and extract 667 ms post-stimulus segments (160 samples) per trial. PhysionetMI contains 69,760 trials across five classes, all used for cross-domain pretraining.  
II-A and II-B each contain 33,000 trials: in II-A, 15,300 (85 characters $\times$ 15 repetitions $\times$ 12 stimuli) form the calibration set and 18,000 (100 characters $\times$ 15 repetitions $\times$ 12 stimuli) form the testing set; all are used for in-domain pretraining. II-B is used for downstream tasks, with 85 calibration characters for fine-tuning and 100 testing characters for evaluation.

\subsection{Experimental Setup}
We evaluate three approaches:  
(1) \textit{Scratch}: Random initialization and supervised training on the calibration set of single-subject P300 data (II-B), serving as the baseline.  
(2) \textit{Cross-domain}: SSL pretraining on motor imagery EEG (PhysionetMI), followed by fine-tuning on the calibration set of single-subject P300 data (II-B).  
(3) \textit{In-domain}: SSL pretraining on in-domain P300 EEG (II-A), followed by fine-tuning on the calibration set of a different subject’s P300 data (II-B).  
P300 aggregation ($G\!\in\!\{1,2,3\}$) is applied only to calibration data (Sec.~\ref{sec:M3}). Evaluation is on the independant II-B test set. The evaluation metrics include Character Recognition Rate (CRR), Accuracy, Binary F1-Score, Fisher’s Discriminant Ratio (FDR), and Information Transfer Rate (ITR), detailed in Appendix~\ref{app:metrics}. All experiments are conducted using 32-bit mixed precision on a single NVIDIA RTX 6000 Ada GPU. 

\subsection{Reconstruction Results}
\begin{figure}
  \centering
  \includegraphics[width=\linewidth]{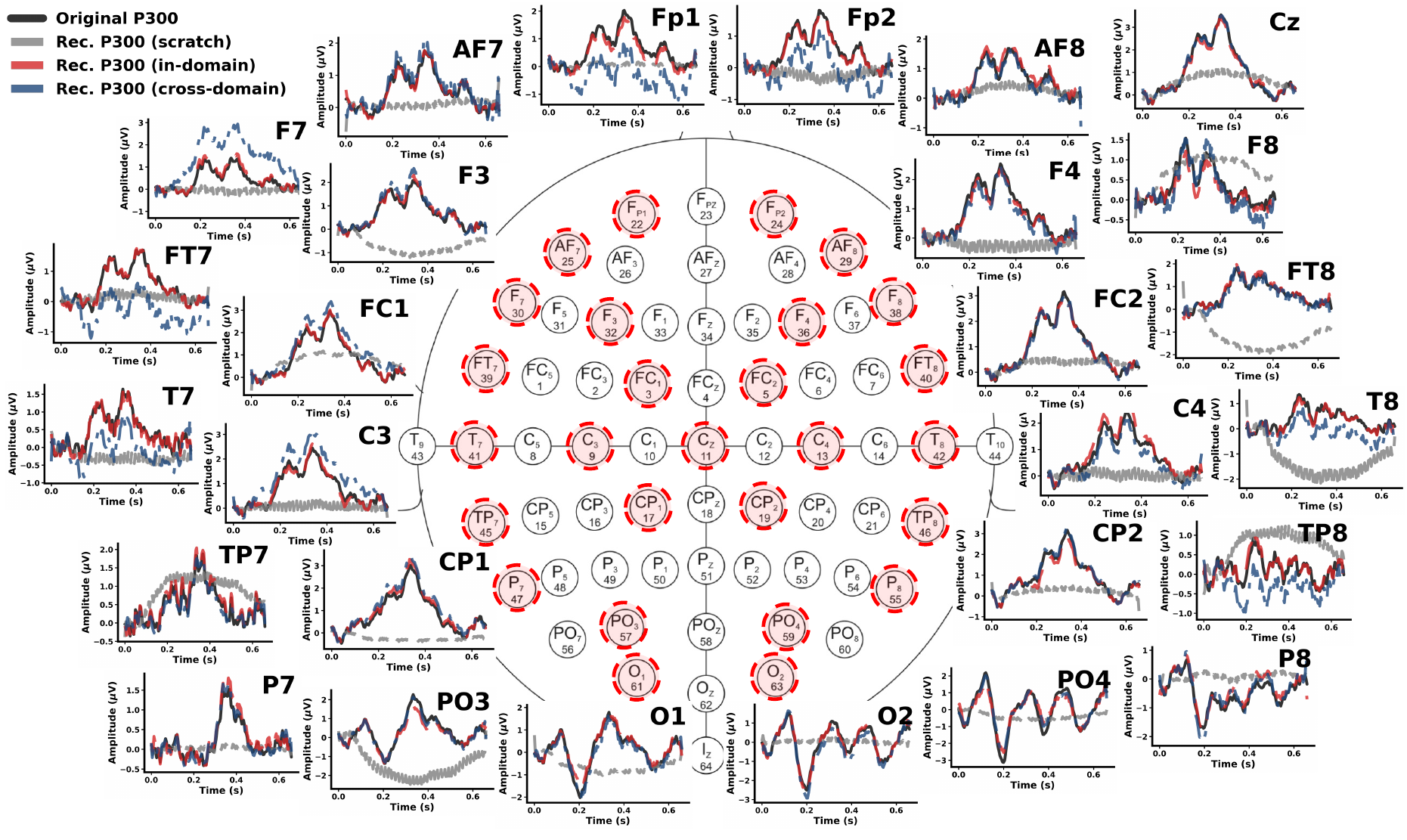}
  \caption{Reconstruction performance across 64 EEG channels. 
  The montage layout is shown in the center, with representative channels displayed around it. 
  Each plot compares the original P300 response (black) with reconstructions from models trained from scratch (gray), with in-domain pretraining (red), and with cross-domain pretraining (blue). More visualizations in Appendix~\ref{app:recon}.}
  \label{fig:3}
\end{figure}

Figure~\ref{fig:3} shows reconstructed EEG signals under the three approaches. 
The ground-truth P300 responses (black) exhibit positive deflections at many frontal and central electrodes (e.g., Fp1, Fp2, Cz), while occipital sites such as O1 and O2 often show negative polarity, likely influenced by the reference montage. 
The model trained from scratch tends to yield noisy reconstructions, whereas in-domain pretraining yields signals closely aligned with the original waveforms. 
Interestingly, cross-domain pretraining on motor imagery data also recovers P300 morphology. For instance, at Cz and AF8 both in-domain (red) and cross-domain (blue) reconstructions match the ground truth, while at F7 and Fp1 the cross-domain model captures the trend but underestimates the amplitude. 

To quantify these observations, we compute the mean squared error (MSE) averaged across all channels and trials. 
The scratch model yields an MSE of 161.38, compared to 3.26 for in-domain pretraining and 7.26 for cross-domain pretraining. 
These results demonstrate that self-supervised pretraining substantially improves EEG reconstruction quality, and that cross-domain pretraining can still learn transferable representations beneficial for downstream P300 tasks.

\subsection{Downstream Results}

\begin{figure}
  \centering
  \begin{subfigure}{0.329\linewidth}
    \centering
    \includegraphics[width=\linewidth]{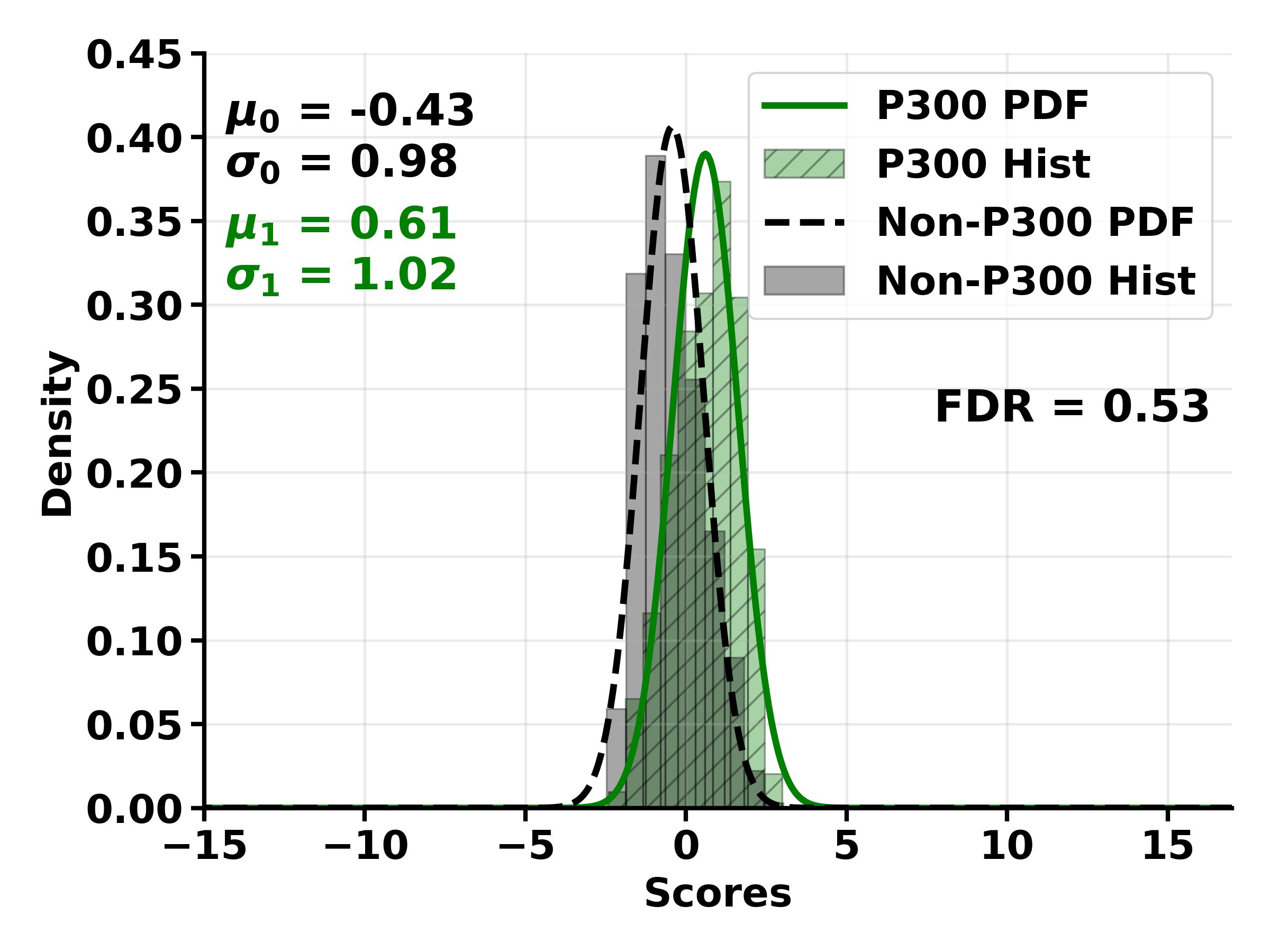}
    \caption{Scratch ($G=1$)}
  \end{subfigure}
  \begin{subfigure}{0.329\linewidth}
    \centering
    \includegraphics[width=\linewidth]{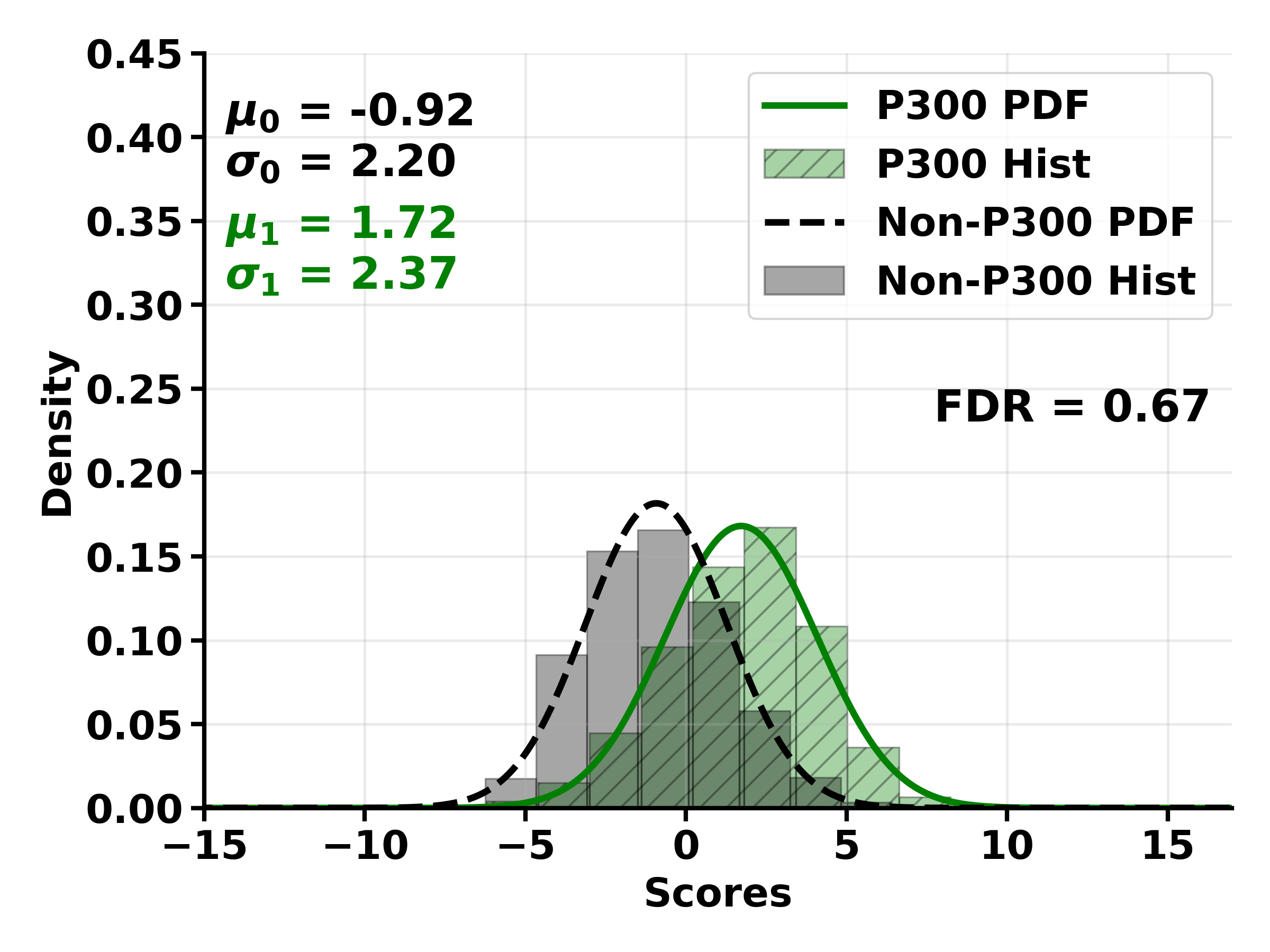}
    \caption{Scratch ($G=2$)}
  \end{subfigure}
  \begin{subfigure}{0.329\linewidth}
    \centering
    \includegraphics[width=\linewidth]{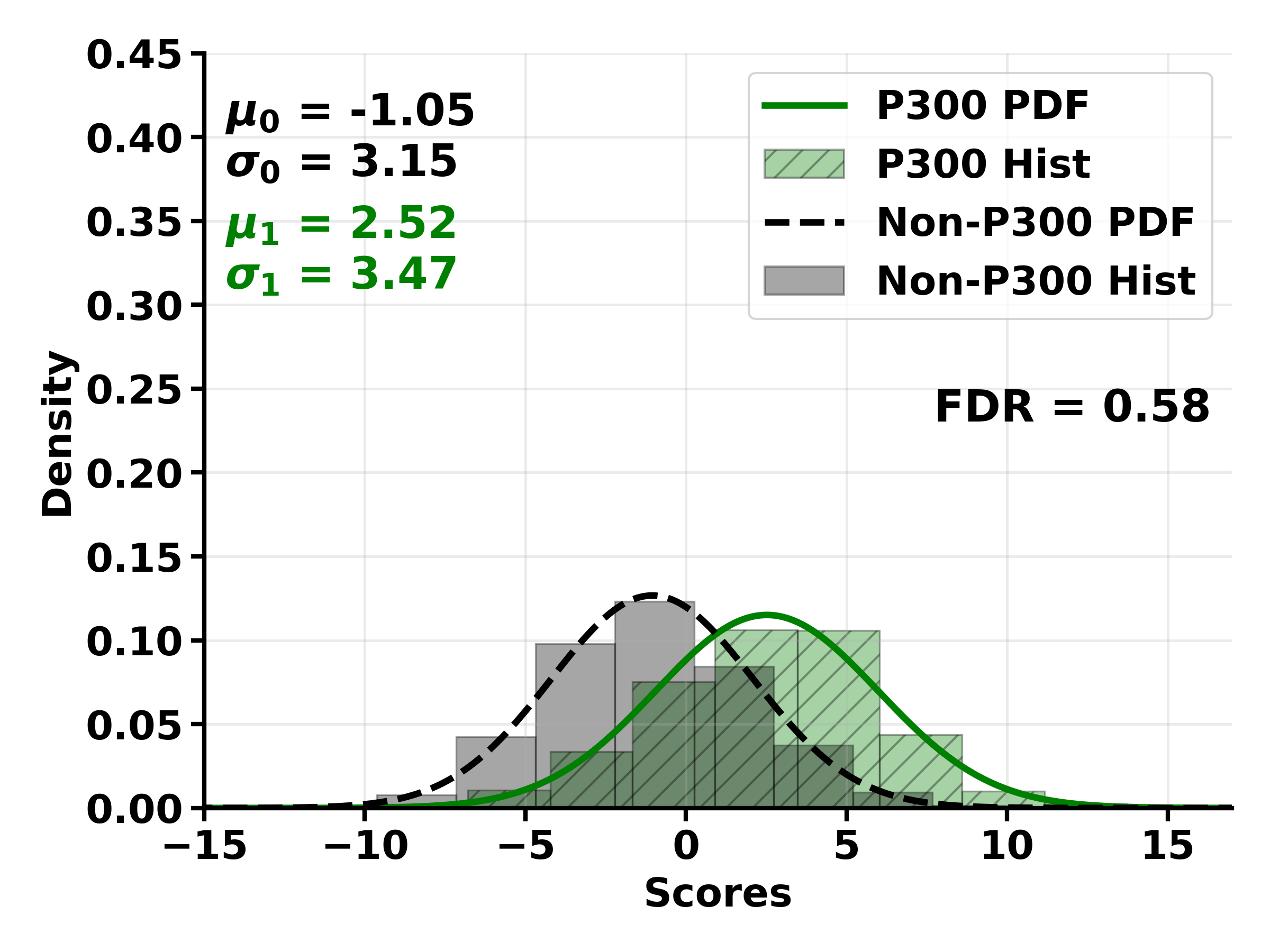}
    \caption{Scratch ($G=3$)}
  \end{subfigure}

  \vspace{0.5cm}
  \begin{subfigure}{0.329\linewidth}
    \centering
    \includegraphics[width=\linewidth]{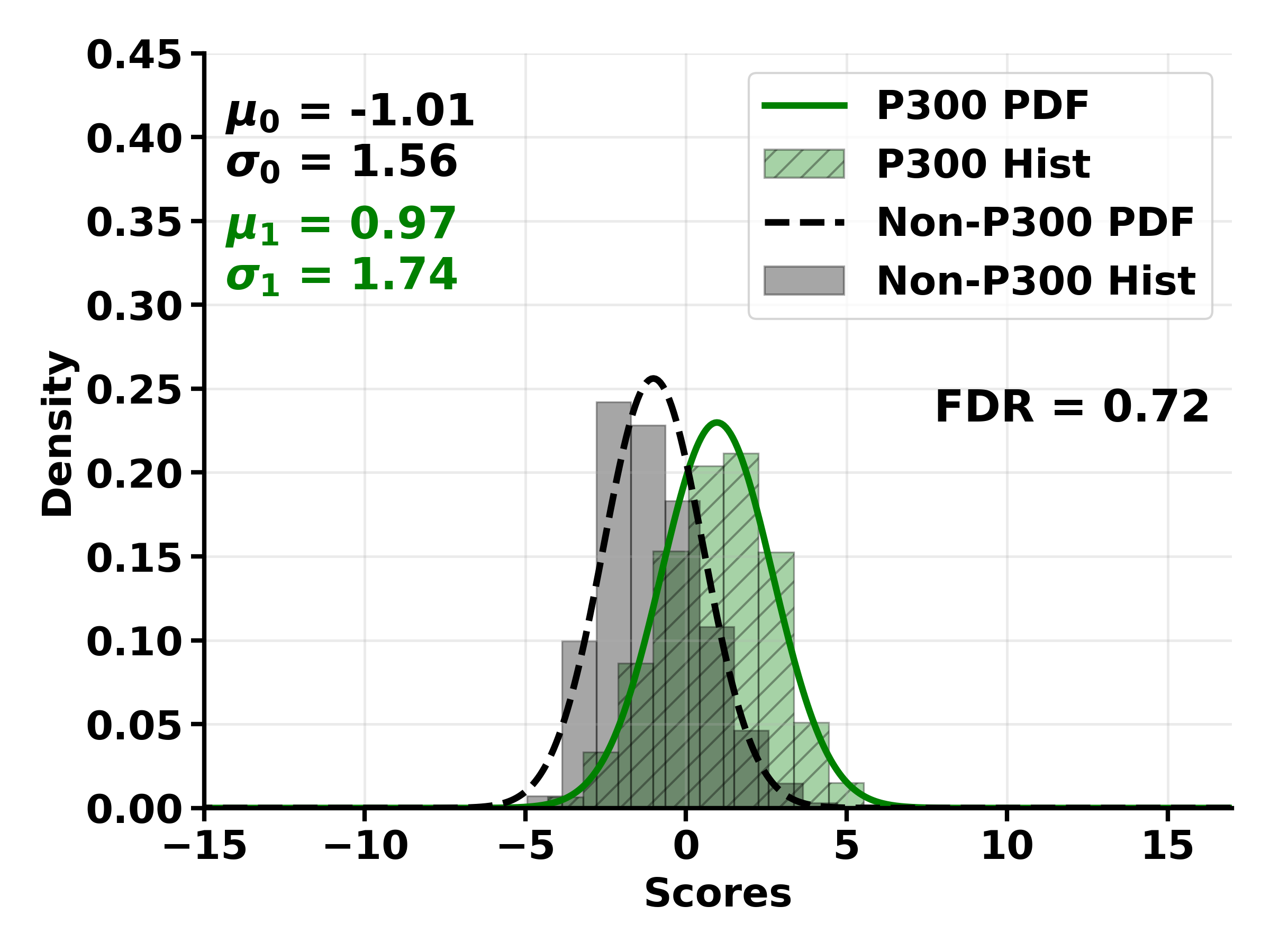}
    \caption{Cross-domain ($G=1$)}
  \end{subfigure}
  \begin{subfigure}{0.329\linewidth}
    \centering
    \includegraphics[width=\linewidth]{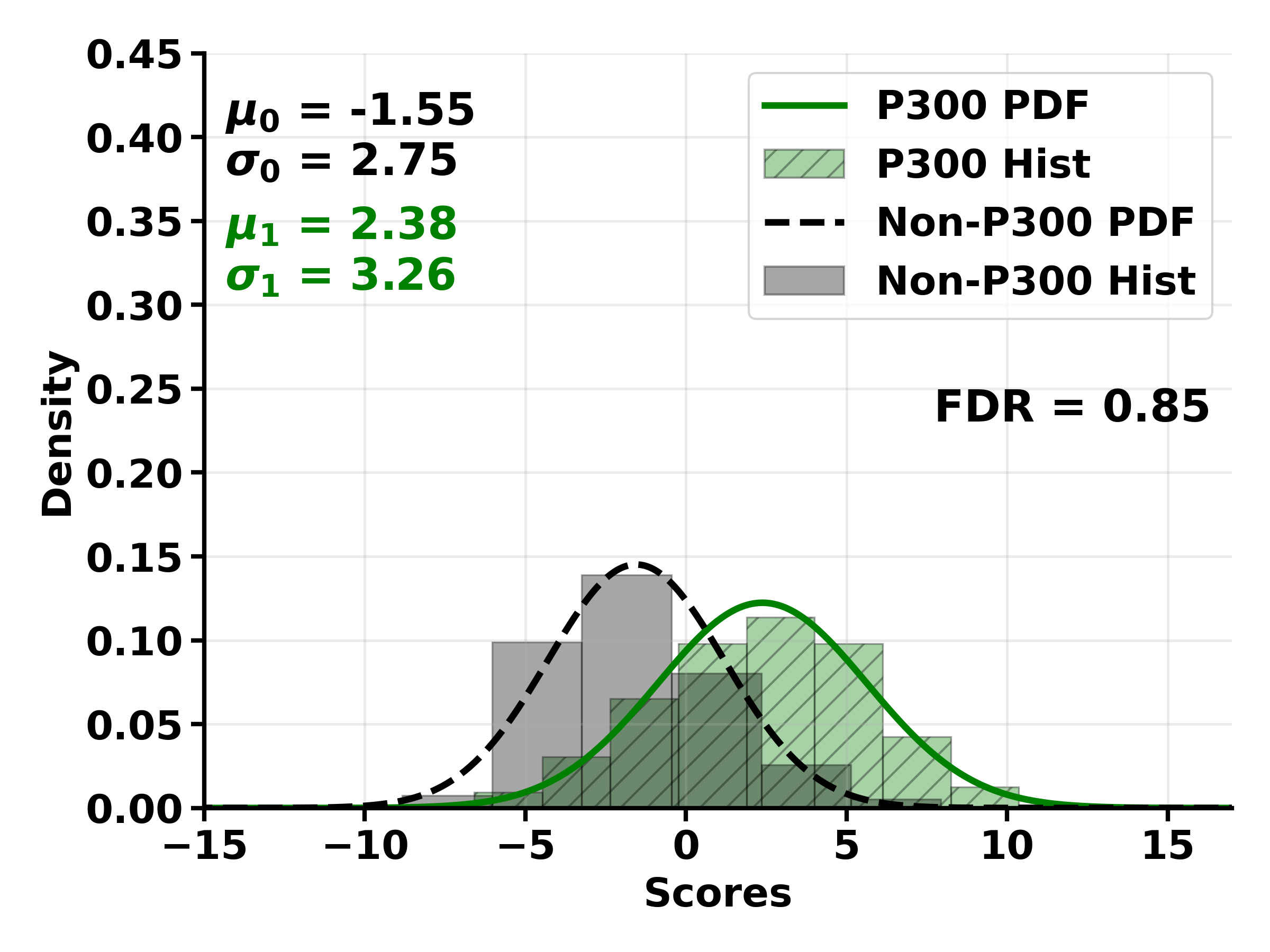}
    \caption{Cross-domain ($G=2$)}
  \end{subfigure}
  \begin{subfigure}{0.329\linewidth}
    \centering
    \includegraphics[width=\linewidth]{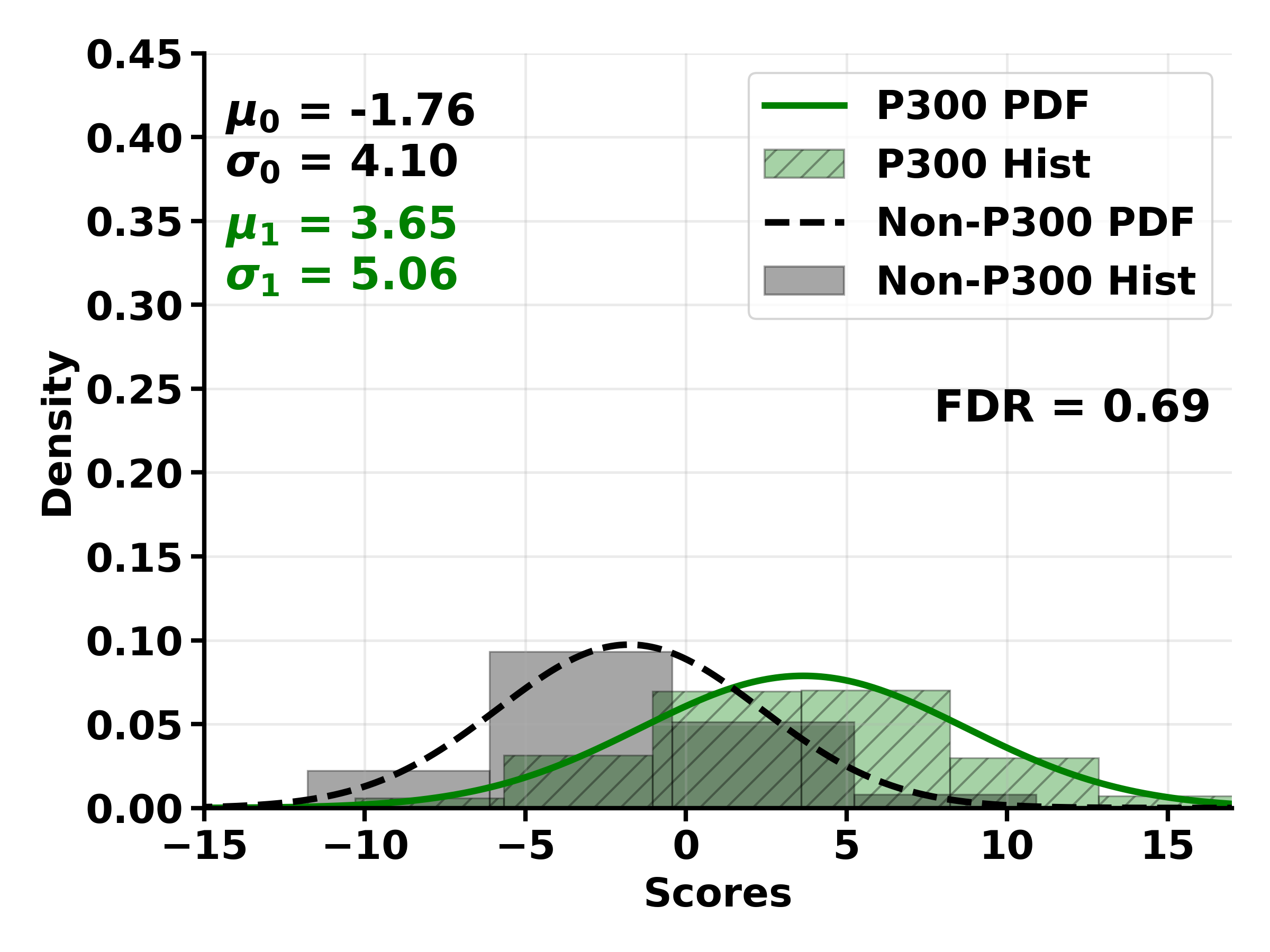}
    \caption{Cross-domain ($G=3$)}
  \end{subfigure}

  \vspace{0.5cm}
  \begin{subfigure}{0.329\linewidth}
    \centering
    \includegraphics[width=\linewidth]{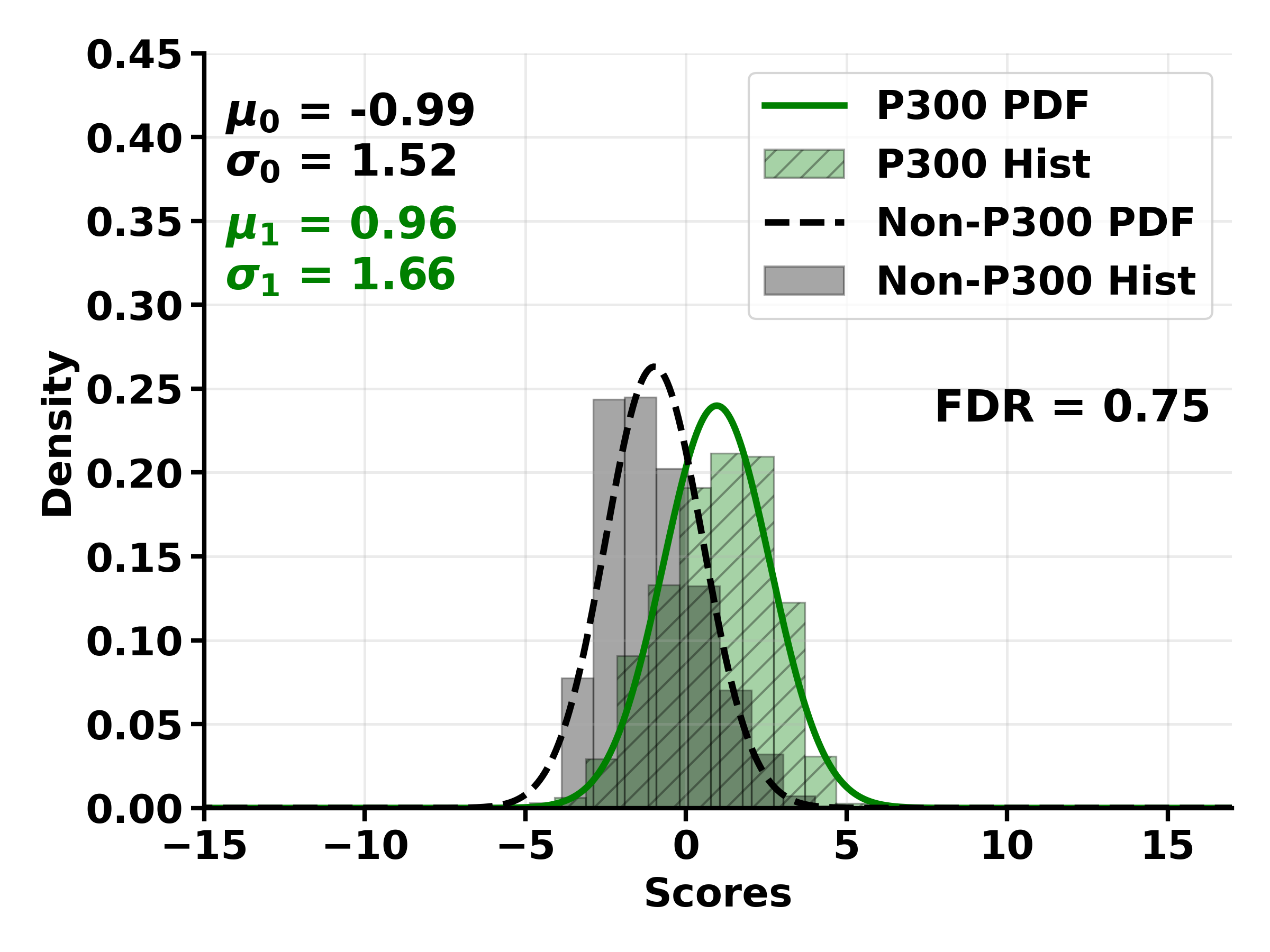}
    \caption{In-domain ($G=1$)}
  \end{subfigure}
  \begin{subfigure}{0.329\linewidth}
    \centering
    \includegraphics[width=\linewidth]{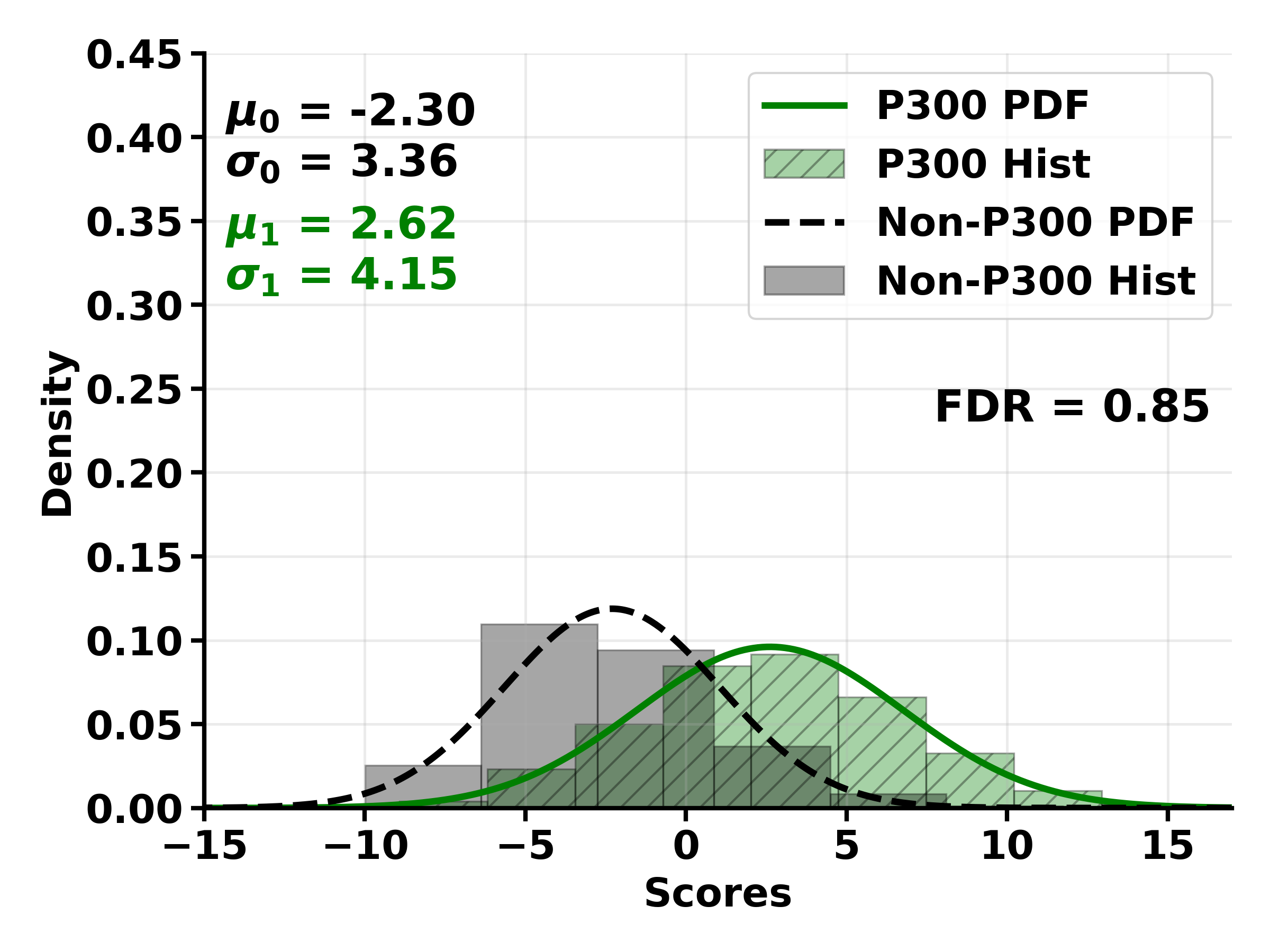}
    \caption{In-domain ($G=2$)}
  \end{subfigure}
  \begin{subfigure}{0.329\linewidth}
    \centering
    \includegraphics[width=\linewidth]{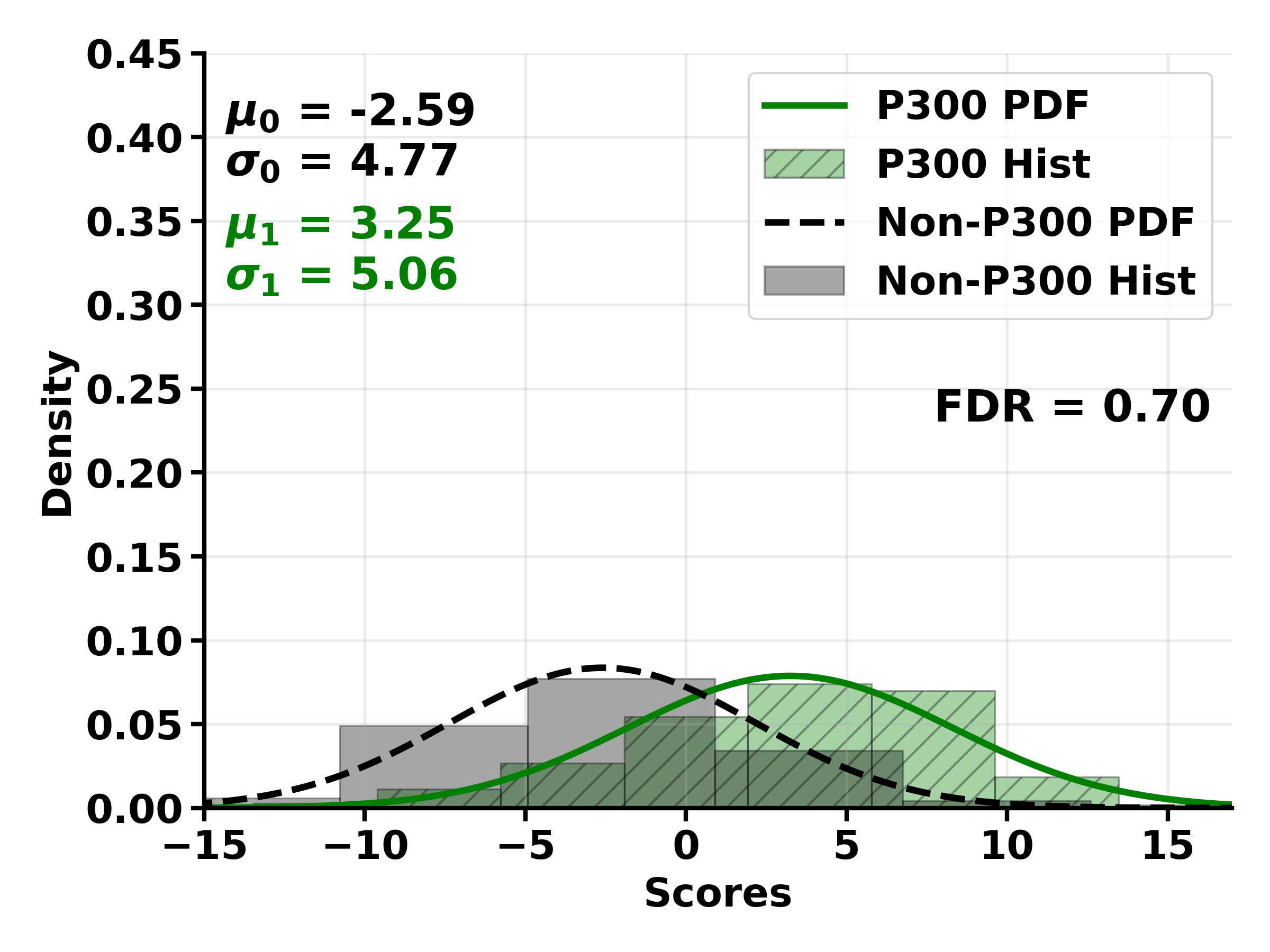}
    \caption{In-domain ($G=3$)}
  \end{subfigure}
  \caption{Distributions of decision scores (log-odds) under three pretraining conditions: 
  (a–c) training from scratch, (d–f) cross-domain pretraining, and (g–i) in-domain pretraining. 
  For each condition, we evaluate three aggregation levels $G\!\in\!\{1,2,3\}$. 
  Green curves and bars show P300 probability density function (PDF) and histogram (Hist), while black dashed curves and gray bars show Non-P300. 
  The Fisher’s Discriminant Ratio (FDR; higher is better) is reported.}
  \label{fig:4}
\end{figure}

Figure~\ref{fig:4} shows decision score distributions for P300 vs. Non-P300 responses under different pretraining and aggregation settings. 
Without aggregation ($G=1$), the two distributions are closer and less separable.
Aggregation ($G=2,3$) generally increases the mean difference between target and non-target responses, thereby improving separability as reflected by higher FDR values. However, when $G=3$, although the mean separation further increases, the variance of both classes also grows, leading to wider and noisier distributions and a drop in FDR compared to $G=2$. For example, in scratch training, FDR improves from 0.53 ($G=1$) to 0.67 ($G=2$), but drops to 0.58 ($G=3$). Overall, in-domain pretraining achieves the highest separability, while cross-domain remains competitive, with both reaching an FDR of 0.85 at $G=2$.

\begin{table}
    \centering
    \caption{Downstream performance under different training and aggregation conditions.}
    \label{tab:1}
    \resizebox{0.95\linewidth}{!}{
    \begin{tabular}{c|c|*{15}{c}|c|c|c}
    \hline
    pretraining & \# \textit{G} & 1  & 2  & 3  & 4  & 5  & 6  & 7  & 8  & 9  & 10 & 11 & 12 & 13 & 14 & 15 & Acc.& B-F1 & FDR\\ \hline
    Scratch   & 1  & 21 & 38 & 49 & 55 & 60 & 67 & 74 & 77 & 83 & 83 & 85 & 87 & 87 & 89 & 89 & 68.58 & 0.4312 & 0.53 \\
    Scratch   & 2  & 31 & 38 & 56 & 65 & 71 & 81 & 83 & 81 & 87 & 89 & \underline{90} & \underline{95} & \underline{93} & \underline{95} & \underline{96} & 68.42 & 0.4481 & 0.67 \\
    Scratch   & 3  & 34 & 43 & 59 & 62 & 69 & 79 & 81 & 84 & 85 & 88 & 88 & 89 & \underline{91} & 89 & \underline{92} & 66.26 & 0.4310 & 0.58 \\ \hline
    Cross-domain   & 1  & 39 & 49 & 56 & 69 & 75 & 81 & 88 & 86 & \underline{90} & 87 & \underline{90} & \underline{93} & \underline{90} & \underline{91} & \underline{92} & 74.00 & 0.4751 & 0.72 \\
    Cross-domain   & 2  & \textbf{44} & \textbf{65} & \textbf{74} & \textbf{75} & 82 & 87 & \underline{93} & \underline{\textbf{92}} & \underline{\textbf{95}} & \underline{95} & \underline{96} & \underline{96} & \underline{96} & \underline{97} & \underline{\textbf{97}} & 73.94 & 0.4937 & \textbf{0.85} \\
    Cross-domain   & 3  & 38 & 56 & 64 & 68 & 78 & 81 & 85 & 84 & \underline{90} & \underline{93}\underline & \underline{95} & \underline{95} & \underline{95} & \underline{97} & \underline{\textbf{97}} & 70.57 & 0.4657 & 0.69 \\ \hline
    In-domain   & 1   & 36 & 49 & 59 & 70 & 72 & 83 & 87 & 84 & 85 & 87 & 86 & \underline{91} & \underline{91} & \underline{92} & \underline{96} & 74.28 & 0.4779 & 0.75\\
    In-domain   & 2 & 43 & \textbf{65} & 69 & 74 & \textbf{84} & \textbf{89} & \underline{\textbf{94}} & \underline{91} & \underline{93} & \underline{\textbf{96}} & \underline{\textbf{98}} & \underline{\textbf{99}} & \underline{\textbf{98}} & \underline{\textbf{99}} & \underline{\textbf{97}} & \textbf{76.18} & 0.\textbf{5074} & \textbf{0.85} \\
    In-domain   & 3 & 40 & 56 & 63 & 70 & 69 & 80 & 87 & 87 & 87 & \underline{91} & \underline{92} & \underline{92} & \underline{92} & \underline{94} & \underline{96} & 71.79 & 0.4689 & 0.70 \\
    \hline
    \end{tabular}
    }
\end{table}

Table~\ref{tab:1} compares character recognition rates (CRR), single-trial accuracy (Acc., \%), binary F1, and FDR under three pretraining conditions across repetitions 1–15. CRRs above 90\% are underlined, and the highest values are highlighted in bold. In-domain pretraining with $G=2$ achieves the best overall performance, reaching 94\% CRR at repetition 7 and the highest single-trial accuracy (76.18\%) and B-F1 (0.5074). Cross-domain pretraining with $G=2$ also performs competitively, reaching 93\% CRR at repetition 7 and FDR = 0.85, demonstrating the transferability of SpellerSSL from motor imagery to P300 tasks and suggesting that pretrained or future foundation models can benefit P300 spellers. Scratch models improve with aggregation but remain consistently worse. Aggregation also accelerates recognition: with $G=2$, both in-domain and cross-domain models surpass 90\% CRR by repetition 7, whereas with $G=1$ this threshold is only reached after repetition 11. Without aggregation and pretraining, the scratch model fails to exceed 90\% CRR even at repetition 15. Excessive aggregation ($G=3$) does not yield further gains, as the increased variance lowers FDR and slightly reduces accuracy. Overall, moderate aggregation ($G=2$) offers the best trade-off, effectively denoising while retaining discriminative ERP features.

\subsection{Calibration Reduction Analysis}
\begin{table*}
    \centering
    \caption{Downstream performance under reduced calibration. 
    }
    \label{tab:2}
    \resizebox{0.95\linewidth}{!}{
    \begin{tabular}{c|c|*{15}{c}|c|c|c}
    \hline
    Calibration & \# \textit{G} & 1  & 2  & 3  & 4  & 5  & 6  & 7  & 8  & 9  & 10 & 11 & 12 & 13 & 14 & 15 & Acc.& B-F1 & FDR\\ \hline
    20\% (Scratch)   & 1  & 3 & 2 & 0 & 4 & 1 & 2 & 3 & 3 & 3 & 4 & 6 & 4 & 4 & 5 & 5 & 59.68 & 0.2556 & 0\\
    20\%   & 1 & 10 & 6 & 17 & 23 & 28 & 28 & 31 & 36 & 38 & 42 & 38 & 36 & 44 & 43 & 46 & 67.24 & 0.3321 & 0.1\\
    20\%   & 2 & 29 & 39 & 45 & 50 & 61 & 68 & 73 & 69 & 70 & 77 & 75 & 82 & 80 & 81 & 81 & 68.11 &  0.4294 & 0.45 \\ \hline
    40\% (Scratch)  & 1 & 10 & 6 & 17 & 23 & 28 & 28 & 31 & 36 & 38 & 42 & 38 & 36 & 44 & 43 & 46 & 67.24 & 0.3321 & 0.1 \\
    40\%  & 1 & 30 & 41 & 49 & 64 & 65 & 74 & 76 & 77 & 82 & 81 & 81 & 89 & 85 & 85 & 86 & 72.50 & 0.4581 & 0.60\\
    40\%  & 2 & 34 & 43 & 51 & 64 & 74 & 76 & 83 & 83 & 86 & 88 & 87 & 88 & 87 & \underline{90} & \underline{92} & 65.37 & 0.4340 & 0.64\\ \hline
    60\% (Scratch) & 1 & 27 & 36 & 50 & 53 & 57 & 64 & 59 & 63 & 66 & 73 & 72 & 77 & 75 & 74 & 77 & 71.22 &0.4265& 0.41\\
    60\%   & 1 & 35 & 43 & 50 & 59 & 66 & 73 & 79 & 79 & 83 & 87 & 88 & \underline{91} & 89 & \underline{90} & \underline{91} & 72.34 & 0.4633 & 0.68\\
    60\%   & 2 & 37 & 59 & 72 & 77 & 82 & 87 & \underline{91} & \underline{92} & \underline{93} & \underline{96} & \underline{96} & \underline{97} & \underline{96} & \underline{97} & \underline{98} & 74.82 & 0.4921 & 0.78\\ \hline
    80\%  (Scratch) & 1 & 23 & 36 & 45 & 56 & 62 & 64 & 69 & 67 & 69 & 76 & 79 & 84 & 81 & 83 & 85 & 68.41 & 0.4292 & 0.47\\
    80\%   & 1 & 37 & 51 & 56 & 64 & 73 & 81 & 86 & 84 & 88 & \underline{91} & \underline{91} & \underline{91} & \underline{92} & \underline{90} & \underline{92} & 73.91 & 0.4754 & 0.70\\
    80\%   & 2 & 43 & 60 & 76 & 78 & 88 & 87 & \underline{91} & \underline{92} & \underline{91} & \underline{96} & \underline{98} & \underline{96} & \underline{94} & \underline{96} & \underline{97} & 71.42 & 0.4761 & 0.81\\ \hline
    100\% (Scratch)   & 1  & 21 & 38 & 49 & 55 & 60 & 67 & 74 & 77 & 83 & 83 & 85 & 87 & 87 & 89 & 89 & 68.58 & 0.4312 & 0.53 \\
    100\%   & 1   & 36 & 49 & 59 & 70 & 72 & 83 & 87 & 84 & 85 & 87 & 86 & \underline{91} & \underline{91} & \underline{92} & \underline{96} & 74.28 & 0.4779 & 0.75\\
    100\%   & 2 & 43 & 65 & 69 & 74 & 84 & 89 & \underline{94} & \underline{91} & \underline{93} & \underline{96} & \underline{98} & \underline{99} & \underline{98} & \underline{99} & \underline{97} & 76.18 & 0.5074 & 0.85 \\

    \hline
    \end{tabular}
    }
\end{table*}

Table~\ref{tab:2} shows the effect of reduced calibration data on downstream performance, reporting results for aggregation $G=1$ and $G=2$, together with single-trial accuracy (Acc.), binary F1 (B-F1), and FDR. Scratch ($G=1$) serves as baseline, and character recognition rates (CRR, \%) above 90\% are underlined. 
Without pretraining and aggregation (Scratch, $G=1$), the model fails to capture discriminative features under limited calibration, yielding FDR $\approx 0$ at 20\%. In-domain pretraining ($G=1$) raises FDR to 0.10, while combining pretraining with aggregation ($G=2$) further boosts FDR to 0.45, enabling CRR to exceed 80\% by repetition 12 even with only 20\% calibration data. As calibration increases, both CRR and FDR improve across all settings, yet the gap between pretrained and scratch models remains substantial, highlighting the advantage of SpellerSSL in low-calibration conditions. At full calibration (100\%), in-domain pretraining with $G=2$ achieves best performance. More broadly, $G=2$ consistently provides the best trade-off: for example, at 60\% calibration, CRR reaches 91\% at repetition 7 with FDR = 0.78, compared to only 0.41 for the scratch baseline.

\begin{figure}
  \centering
  \begin{subfigure}{0.49\linewidth}
    \centering
    \includegraphics[width=\linewidth]{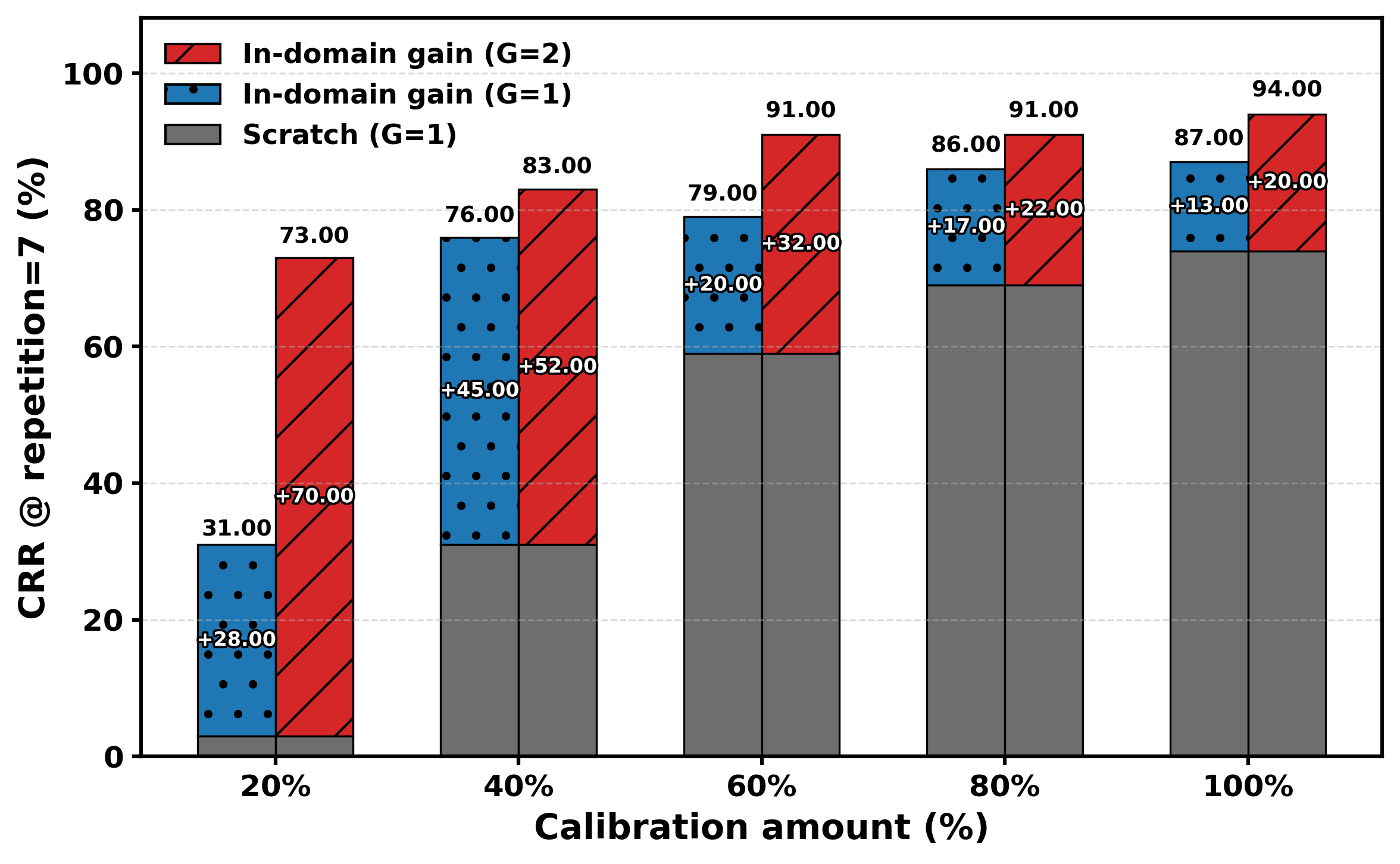}
    \caption{Character recognition rate (CRR) at repetition 7.}
    \label{fig:5a}
  \end{subfigure}
  \begin{subfigure}{0.49\linewidth}
    \centering
    \includegraphics[width=\linewidth]{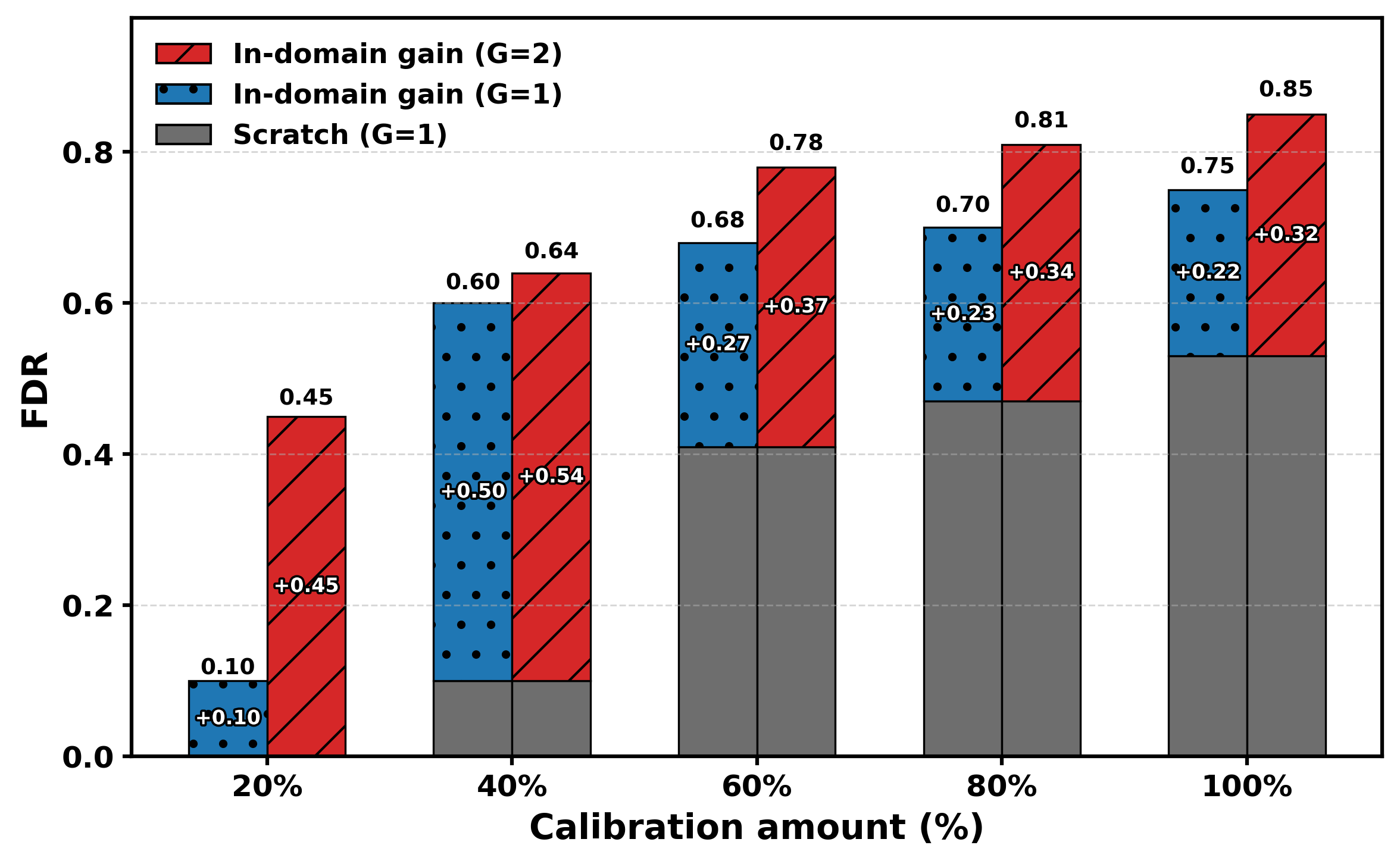}
    \caption{Fisher’s Discriminant Ratio (FDR). }
    \end{subfigure}
    \label{fig:5b}
    \caption{Calibration reduction analysis using in-domain checkpoints. Baseline results from scratch ($G=1$) are shown in gray, while gains from in-domain pretraining with $G=1$ and $G=2$ are shown in blue and red, respectively.}
  \label{fig:5}
\end{figure}

Figure~\ref{fig:5} visualizes the improvements of in-domain pretraining and P300 aggregation over the scratch baseline ($G=1$ without aggregation). (a) shows CRR at repetition 7, where the earliest time CRR can exceed 90\%. (b) presents the corresponding FDR values. Across all calibration levels, SpellerSSL with in-domain pretraining substantially improves early recognition and increases FDR, while combining pretraining with $G=2$ aggregation further enhances performance. Overall, these results highlight that self-supervised pretraining reduces calibration requirements, and that moderate P300 aggregation ($G=2$) provides additional gains, improving both recognition speed and discriminability in the P300 speller task.

\subsection{Baseline Comparision}
\begin{table}
\caption{Comparison of ITRs (bits/min) between SpellerSSL and state-of-the-art models.}\label{tab:itr}
\large
\resizebox{\linewidth}{!}{
\begin{tabular}{lc|ccccccccccccccc|c}
\hline
\multicolumn{1}{l|}{Model}          & Ref.    & 1           & 2           & 3           & 4           & 5           & 6           & 7           & 8           & 9           & 10          & 11          & 12          & 13          & 14          & 15   & Ave.      \\ \hline
\multicolumn{1}{l|}{ST-CapsNet}     &   \cite{wang2023st}      & 15.22 & 19.74 & 17.05 & \textbf{18.06} & \textbf{17.5} & 15.37 & 15.2 & 13.02 & 11.98 & 11.81 & 11.07 & 10.23 & 9.32 & 9.07 & 8.33 & 13.53 \\
\multicolumn{1}{l|}{ERP-CapsNet}    &    \cite{ma2021capsule}     & 17.69 & 19.23 & 17.05 & 16.2 & 16.13 & 15.06 & 13.76 & 13.02 & 11.74 & 11.58 & 10.63 & 9.82 & 9.13 & 8.7 & 8.33 & 13.2    \\
\multicolumn{1}{l|}{CNN-1}          &    \cite{cecotti2010convolutional}     & 11.76 & 15.3 & 14.25 & 14.44 & 15.47 & 13.88 & 12.44 & 12.76 & 12.22 & 10.91 & 10.01 & 9.07 & 8.6 & 8.2 & 7.69 & 11.8      \\
\multicolumn{1}{l|}{MCNN-1}         &   \cite{cecotti2010convolutional}      & 14.04 & 16.74 & 15.43 & 13.1 & 14.83 & 13.32 & 13.49 & 13.55 & 11.98 & 11.13 & 10.84 & 10.02 & 9.32 & 8.53 & 8.0 & 12.29     \\
\multicolumn{1}{l|}{S-LDA}          &   \cite{blankertz2011single, ma2021capsule}      & 14.04 & 16.25 & 16.23 & 14.1 & 14.2 & 13.04 & 13.22 & 12.51 & 11.06 & 11.13 & 10.21 & 10.23 & 8.95 & 8.53 & 8.33 & 12.14      \\
\multicolumn{1}{l|}{SWFP}           &    \cite{gerson2006cortically, wang2023st}     & 14.63 & 19.23 & 18.32 & 15.49 & 15.79 & 14.17 & 13.49 & 12.76 & 11.51 & 11.13 & 10.21 & 9.63 & 8.77 & 8.53 & 8.0 & 12.78        \\
\multicolumn{1}{l|}{P3T}            &    \cite{hong2024p3t}     & 15.83 & 20.26 & \textbf{20.52} & 15.84 & 16.13 & 15.06 & 14.61 & 13.28 & 12.22 & 11.13 & 10.63 & 9.82 & 9.13 & 8.53 & 8.33 & 13.42    \\ \hline
\multicolumn{2}{l|}{Cross-domain )}     & 17.06 & \textbf{21.86} & \textbf{20.52} & 16.93 & 16.46 & 15.68 & 15.51 & \textbf{13.55} & \textbf{12.97} & 11.81 & 11.07 & 10.23 & 9.51 & 9.07 & 8.51 & \textbf{14.05} \\
\multicolumn{2}{l|}{In-domain }        & 16.44 & \textbf{21.86} & 18.32 & 16.57 & 17.15 & \textbf{16.31} & \textbf{15.82} & 13.28 & 12.46 & 12.06 & \textbf{11.55} & \textbf{10.91} & \textbf{9.92} & \textbf{9.48}& 8.51 & 14.04 \\
\multicolumn{2}{l|}{In-domain (60\%)}  &12.88 & 18.72 & 19.62 & 17.68 & 16.46 & 15.68 & 14.9 & 13.55 & 12.46 & \textbf{12.06} & 11.07 & 10.44 & 9.51 & 9.07 & \textbf{8.69} & 13.52 \\ \hline
\end{tabular}
}
\end{table}
Table~\ref{tab:itr} compares SpellerSSL ($G=2$) with state-of-the-art (SOTA) methods in terms of information transfer rate (ITR, bits/min). 
Details of ITR computation are provided in Appendix~\ref{app:itr}. The best results are highlighted in bold, and average ITRs across repetitions are also reported to capture overall performance. The baselines include CNN-1~\cite{cecotti2010convolutional}, MCNN-1~\cite{cecotti2010convolutional}, ERP-CapsNet~\cite{ma2021capsule}, ST-CapsNet~\cite{wang2023st}, Spatially Weighted FLD-PCA (SWFP)~\cite{gerson2006cortically, wang2023st}, Shrinkage LDA (S-LDA)~\cite{blankertz2011single, ma2021capsule}, and stacked Transformers (P3T)~\cite{hong2024p3t}. 
Overall, SpellerSSL achieves superior or comparable results in ITR.  The peak ITR of 21.86 bits/min is obtained at repetition 2 by both cross-domain and in-domain SpellerSSL, surpassing all SOTA baselines. Cross-domain pretraining further achieves the highest average ITR (14.05 bits/min), slightly above in-domain (14.04 bits/min), highlighting the transferability of self-supervised pretraining even across paradigms. 
A complementary comparison of CRRs (\%) across repetitions with SOTA models is provided in Appendix~\ref{app:baseline_crr}.

\section{Conclusion}
We proposed SpellerSSL, a self-supervised framework for P300 speller BCIs that combines reconstruction-based pretraining with a lightweight ERP-Head classifier, together with a P300 aggregation strategy to improve ERP signal quality. Experiment results show that SpellerSSL outperforms or matches state-of-the-art methods in information transfer rate (ITR). In-domain SSL with moderate aggregation achieves the best overall results, reducing calibration requirements by up to 60\% while maintaining the performance of prior models. Cross-domain SSL further demonstrates strong transferability, highlighting the potential of self-supervised EEG pretraining. These results establish SSL and aggregation as effective strategies for improving efficiency, robustness, and generalization in practical P300 speller BCIs, paving the way toward future EEG foundation models.

\medskip
{
\small
\bibliographystyle{unsrt}
\bibliography{reference}

@article{lotte2018review,
  title={A review of classification algorithms for {EEG}-based brain--computer interfaces: a 10 year update},
  author={Lotte, Fabien and Bougrain, Laurent and Cichocki, Andrzej and Clerc, Maureen and Congedo, Marco and Rakotomamonjy, Alain and Yger, Florian},
  journal={Journal of Neural Engineering},
  volume={15},
  number={3},
  pages={031005},
  year={2018},
  publisher={iOP Publishing}
}

@article{bacsar1984new,
  title={A new approach to endogenous event-related potentials in man: relation between {EEG} and {P300}-wave},
  author={Ba{\c{s}}ar, E and Ba{\c{s}}ar-Eroglu, C and Rosen, B and Sch{\"u}tt, A},
  journal={International Journal of Neuroscience},
  volume={24},
  number={1},
  pages={1--21},
  year={1984},
  publisher={Taylor \& Francis}
}

@INPROCEEDINGS{hong2024p3t,
  author={Hong, Jiazhen and Najafizadeh, Laleh},
  booktitle={2024 58th Asilomar Conference on Signals, Systems, and Computers}, 
  title={{P3T}: A Transformer Model for Enhancing Character Recognition Rates in {P300} Speller Systems}, 
  year={2024},
  volume={},
  number={},
  pages={514-518},
  doi={10.1109/IEEECONF60004.2024.10942979}}

@article{cecotti2010convolutional,
  title={Convolutional neural networks for {P300} detection with application to brain--computer interfaces},
  author={Cecotti, Hubert and Graser, Axel},
  journal={IEEE Transactions on Pattern Analysis and Machine Intelligence},
  volume={33},
  number={3},
  pages={433--445},
  year={2010}
}

@article{ma2021capsule,
  title={Capsule network for {ERP} detection in brain--computer interface},
  author={Ma, Ronghua and Yu, Tianyou and Zhong, Xiaoli and Yu, Zhu Liang and Li, Yuanqing and Gu, Zhenghui},
  journal={IEEE Transactions on Neural Systems and Rehabilitation Engineering},
  volume={29},
  pages={718--730},
  year={2021}
}

@article{wang2023st,
  title={{ST}-{C}aps{N}et: Linking Spatial and Temporal Attention with Capsule Network for {P300} Detection Improvement},
  author={Wang, Zehui and Chen, Chuangquan and Li, Junhua and Wan, Feng and Sun, Yu and Wang, Hongtao},
  journal={IEEE Transactions on Neural Systems and Rehabilitation Engineering},
  volume={31},
  pages={991--1000},
  year={2023}
}

@inproceedings{hong2025topoeeg,
  title={{TopoEEG}: A Timesformer-Based Topographic Image Representation Method for Single-Trial Early Detection of {P300}},
  author={Hong, Jiazhen and Najafizadeh, Laleh},
  booktitle={2025 IEEE 22nd International Symposium on Biomedical Imaging (ISBI)},
  pages={1--4},
  year={2025},
  organization={IEEE}
}

@inproceedings{akram2013novel,
  title={A novel {P300}-based {BCI} system for words typing},
  author={Akram, Faraz and Metwally, Mohamed K and Han, Hee-Sok and Jeon, Hyun-Jae and Kim, Tae-Seong},
  booktitle={2013 International Winter Workshop on Brain-Computer Interface (BCI)},
  pages={24--25},
  year={2013},
  organization={IEEE}
}

@article{hong2024chatbci,
  title={{ChatBCI}: A {P300} Speller {BCI} Leveraging Large Language Models for Improved Sentence Composition in Realistic Scenarios},
  author={Hong, Jiazhen and Wang, Weinan and Najafizadeh, Laleh},
  journal={preprint arXiv:2411.15395},
  year={2024}
}

@article{speier2018improving,
  author = {Speier, William and Arnold, Corey and Chandravadia, Nand and Roberts, Dustin and Pendekanti, Shrita and Pouratian, Nader},
  title = {Improving {P300} spelling rate using language models and predictive spelling},
  journal = {Brain-Computer Interfaces},
  year = {2018},
  volume = {5},
  number = {1},
  pages = {13--22},
  publisher = {Taylor \& Francis}
}

@inproceedings{ronneberger2015unet,
  title={U-net: Convolutional networks for biomedical image segmentation},
  author={Ronneberger, Olaf and Fischer, Philipp and Brox, Thomas},
  booktitle={Medical image computing and computer-assisted intervention--MICCAI 2015: 18th international conference, Munich, Germany, October 5-9, 2015, proceedings, part III 18},
  pages={234--241},
  year={2015},
  organization={Springer}
}

@inproceedings{smith2019super,
  title={Super-convergence: Very fast training of neural networks using large learning rates},
  author={Smith, Leslie N and Topin, Nicholay},
  booktitle={Artificial intelligence and machine learning for multi-domain operations applications},
  volume={11006},
  pages={369--386},
  year={2019},
  organization={SPIE}
}

@article{chandravadia2025comparing,
  title={Comparing {P300} flashing paradigms in online typing with language models},
  author={Chandravadia, Nand and Pendekanti, Shrita and Roberts, Dustin and Tran, Robert and Panchavati, Saarang and Arnold, Corey and Pouratian, Nader and Speier, William},
  journal={PlOS ONE},
  volume={20},
  number={2},
  pages={e0303390},
  year={2025},
  publisher={Public Library of Science San Francisco, CA USA}
}

@techreport{krusienski2004bci,
  author       = {Krusienski, Dean and Schalk, Gerwin},
  title        = {{BCI} Competition {III} Challenge},
  year         = {2004},
  note         = {URL: https://www.bbci.de/competition/iii/desc\_II.pdf}
}

@article{moabb2018,
  title={{MOABB}: trustworthy algorithm benchmarking for {BCIs}},
  author={Jayaram, Vinay and Barachant, Alexandre},
  journal={Journal of Neural Engineering},
  volume={15},
  number={6},
  pages={066011},
  year={2018},
  publisher={IOP Publishing}
}

@article{schalk2004bci2000,
  title={{BCI2000}: a general-purpose brain-computer interface ({BCI}) system},
  author={Schalk, Gerwin and McFarland, Dennis J and Hinterberger, Thilo and Birbaumer, Niels and Wolpaw, Jonathan R},
  journal={IEEE Transactions on Biomedical Engineering},
  volume={51},
  number={6},
  pages={1034--1043},
  year={2004},
  publisher={IEEE}
}

@misc{goldberger2000eegmmidb,
  author       = {Goldberger, Ary L. and Amaral, Luis A. Nunes and Glass, Leon and Hausdorff, Jeffrey M. and Ivanov, Plamen Ch. and Mark, Roger G. and Mietus, Joseph E. and Moody, George B. and Peng, Chung-Kang and Stanley, H. Eugene},
  title        = {{EEG} Motor Movement/Imagery Dataset},
  year         = {2009},
  howpublished = {\url{https://physionet.org/content/eegmmidb/1.0.0/}},
  note         = {PhysioNet},
}

@article{blankertz2011single,
  title={Single-trial analysis and classification of {ERP} components—a tutorial},
  author={Blankertz, Benjamin and Lemm, Steven and Treder, Matthias and Haufe, Stefan and M{\"u}ller, Klaus-Robert},
  journal={NeuroImage},
  volume={56},
  number={2},
  pages={814--825},
  year={2011},
  publisher={Elsevier}
}

@article{gerson2006cortically,
  title={Cortically coupled computer vision for rapid image search},
  author={Gerson, Adam D and Parra, Lucas C and Sajda, Paul},
  journal={IEEE Transactions on Neural Systems and Rehabilitation Engineering},
  volume={14},
  number={2},
  pages={174--179},
  year={2006},
  publisher={IEEE}
}
}

\appendix
\section{Model Architectures}

\subsection{1D U-Net Backbone}
\label{app:unet}
Table~\ref{tab:unet1d_backbone} summarizes the customized 1D U-Net used for SSL pretraining. 
During downstream adaptation, we only use the \textit{bottleneck} (encoder bottleneck, pre-decoder) feature map for classification, whereas the pretraining objective is applied over the entire reconstruction pipeline (encoder + bottleneck + decoder). 
Input sequences are zero-padded so that \(L \bmod 16 = 0\).

\begin{table}[h]
\centering
\caption{1D U-Net backbone used for SSL pretraining. The feature map extracted for downstream is at the bottleneck (highlighted in bold).}
\label{tab:unet1d_backbone}
\small
\begin{tabular}{lcccc}
\hline
\textbf{Stage} & \textbf{Channels ($\text{in}\rightarrow\text{out}$)} & \textbf{Kernel} & \textbf{Stride} & \textbf{Output length} \\
\hline
Enc1 Conv$\times$2        & $C \rightarrow 64$    & $3$ & $1$ & $L$ \\
MaxPool                   & $64 \rightarrow 64$   & $2$ & $2$ & $L/2$ \\
Enc2 Conv$\times$2        & $64 \rightarrow 128$  & $3$ & $1$ & $L/2$ \\
MaxPool                   & $128 \rightarrow 128$ & $2$ & $2$ & $L/4$ \\
Enc3 Conv$\times$2        & $128 \rightarrow 256$ & $3$ & $1$ & $L/4$ \\
MaxPool                   & $256 \rightarrow 256$ & $2$ & $2$ & $L/8$ \\
Enc4 Conv$\times$2        & $256 \rightarrow 512$ & $3$ & $1$ & $L/8$ \\
MaxPool                   & $512 \rightarrow 512$ & $2$ & $2$ & $L/16$ \\
\textbf{Bottleneck Conv$\times$2} & $512 \rightarrow 1024$ & $3$ & $1$ & $L/16$ \\
UpConv + Dec4 Conv$\times$2 & $1024 \rightarrow 512$ & $3$ & $2$ & $L/8$ \\
UpConv + Dec3 Conv$\times$2 & $512 \rightarrow 256$  & $3$ & $2$ & $L/4$ \\
UpConv + Dec2 Conv$\times$2 & $256 \rightarrow 128$  & $3$ & $2$ & $L/2$ \\
UpConv + Dec1 Conv$\times$2 & $128 \rightarrow 64$   & $3$ & $2$ & $L$ \\
Final Conv$1\times1$        & $64 \rightarrow C$    & $1$ & $1$ & $L$ \\
\hline
\end{tabular}
\end{table}

\subsection{ERP-Head Classifier}\label{app:erphead}
The ERP-Head operates on the encoder bottleneck \emph{sequence}
\[
b \in \mathbb{R}^{1024 \times \tfrac{L}{16}},
\]
discarding the decoder. It applies a pointwise projection, followed by
a depthwise temporal convolution, a dilated depthwise convolution, and a final pointwise fusion.
Global average pooling (GAP) over time yields a feature vector
\(h \in \mathbb{R}^{D}\), which is then passed to a fully connected layer for binary classification.

\begin{table}[h]
\centering
\caption{ERP-Head (tconv) used for downstream P300 classification.}
\label{tab:tconv_head}
\small
\begin{tabular}{lcccc}
\hline
\textbf{Layer} & \textbf{Channels ($\text{in}\!\rightarrow\!\text{out}$)} & \textbf{Kernel} & \textbf{Stride / Dilation} & \textbf{Output length} \\
\hline
Conv $1\times1$ (projection) & $1024 \rightarrow D$ & $1$ & $1$ / -- & $L/16$ \\
Depthwise Conv               & $D \rightarrow D$    & $3$ & $1$ / $1$ & $L/16$ \\
Dilated Depthwise Conv       & $D \rightarrow D$    & $3$ & $1$ / $2$ & $L/16$ \\
Conv $1\times1$ (fusion)     & $D \rightarrow D$    & $1$ & $1$ / -- & $L/16$ \\
Global Avg. Pool (time)      & $D \rightarrow D$    & --  & --        & $1$ \\
Fully Connected (classifier) & $D \rightarrow 2$    & --  & --        & $1$ \\
\hline
\end{tabular}
\end{table}
\vspace{2pt}
\footnotesize\emph{Notes.}
All convolutions use GELU activations; the first $1{\times}1$ projection is followed by BatchNorm1d.
All convolutions use stride $=1$ with padding chosen to preserve the temporal length (hence $L/16$ before GAP).
“Output length” refers to the temporal dimension; the final fully connected layer outputs $2$ logits (non-target vs. target) while the temporal length is $1$ after GAP.
\normalsize

\section{Evaluation Metrics}\label{app:metrics}
Each character selection uses $R{=}15$ repetitions, with 12 flashes (6 rows and 6 columns) per repetition.  
Let $s_k^{(i)}$ be the classifier score for code $k$ in repetition $i$. The cumulative score up to $n$ repetitions is
\begin{equation}
S_k(n) = \sum_{i=1}^n s_k^{(i)}.
\end{equation}
The predicted column $c$ and row $r$ are
\begin{equation}
c = \arg\max_{k\le 6} S_k(n), \qquad
r = \arg\max_{k>6} S_k(n),
\end{equation}
and the recognized character is given by $(r,c)$.

\subsection{Character Recognition Rate (CRR)}
Proportion of correctly recognized characters after $n$ repetitions.

\subsection{Single-trial Accuracy}
Binary classification accuracy at the trial level, with labels
\[
y[k]=
\begin{cases}
1,& k\in\mathcal{K}_{\mathrm{P300}} \ (\text{P300}),\\
0,& \text{otherwise} \ (\text{Non-P300}).
\end{cases}
\]

\subsection{Binary F1-Score}
For trial-level P300 vs.~Non-P300 classification, the F1-score balances precision and recall:
\begin{equation}
\text{F1} = \frac{2 \cdot \text{Precision} \cdot \text{Recall}}
                 {\text{Precision} + \text{Recall}},
\quad
\text{Precision}=\tfrac{TP}{TP+FP},\ 
\text{Recall}=\tfrac{TP}{TP+FN}.
\end{equation}

\subsection{Fisher's Discriminant Ratio (FDR)}
\begin{equation}
\text{FDR} 
= \frac{(\mu_{\text{P300}} - \mu_{\text{Non}})^{2}}
       {\sigma_{\text{P300}}^{2} + \sigma_{\text{Non}}^{2}}
\end{equation}
where $\mu$ and $\sigma$ denote the mean and variance of scores for P300 vs.~Non-P300 trials, respectively.

\subsection{Information Transfer Rate (ITR)} \label{app:itr}
Information transfer rate (ITR) quantifies communication efficiency in bits/min, balancing accuracy and speed~\cite{wang2023st}:
\begin{equation}
\mathrm{ITR}(r) 
= \tfrac{60}{T(r)} \left[
\log_2 N
+ A(r)\log_2 A(r) 
+ (1-A(r))\log_2\tfrac{1-A(r)}{N-1}
\right],
\end{equation}
where $A(r)$ is the CRR after $r$ repetitions, $N$ is the number of characters, and $T(r)$ is the average time per selection.

In this study, we follow the BCI Competition III-II dataset~\cite{krusienski2004bci}. 
Each repetition includes 12 flashes, each lasting $100$~ms with a $75$~ms inter-stimulus interval, totaling $2.1$~s. 
Together with a fixed $2.5$~s pause between characters, the cumulative time after $r$ repetitions is
\begin{equation}\label{eq:t}
    T(r) = 2.5 + 2.1 \times r.
\end{equation}

\section{Additional ERP Visualizations}\label{app:recon}
\begin{figure}
  \centering
  \includegraphics[width=0.9\linewidth]{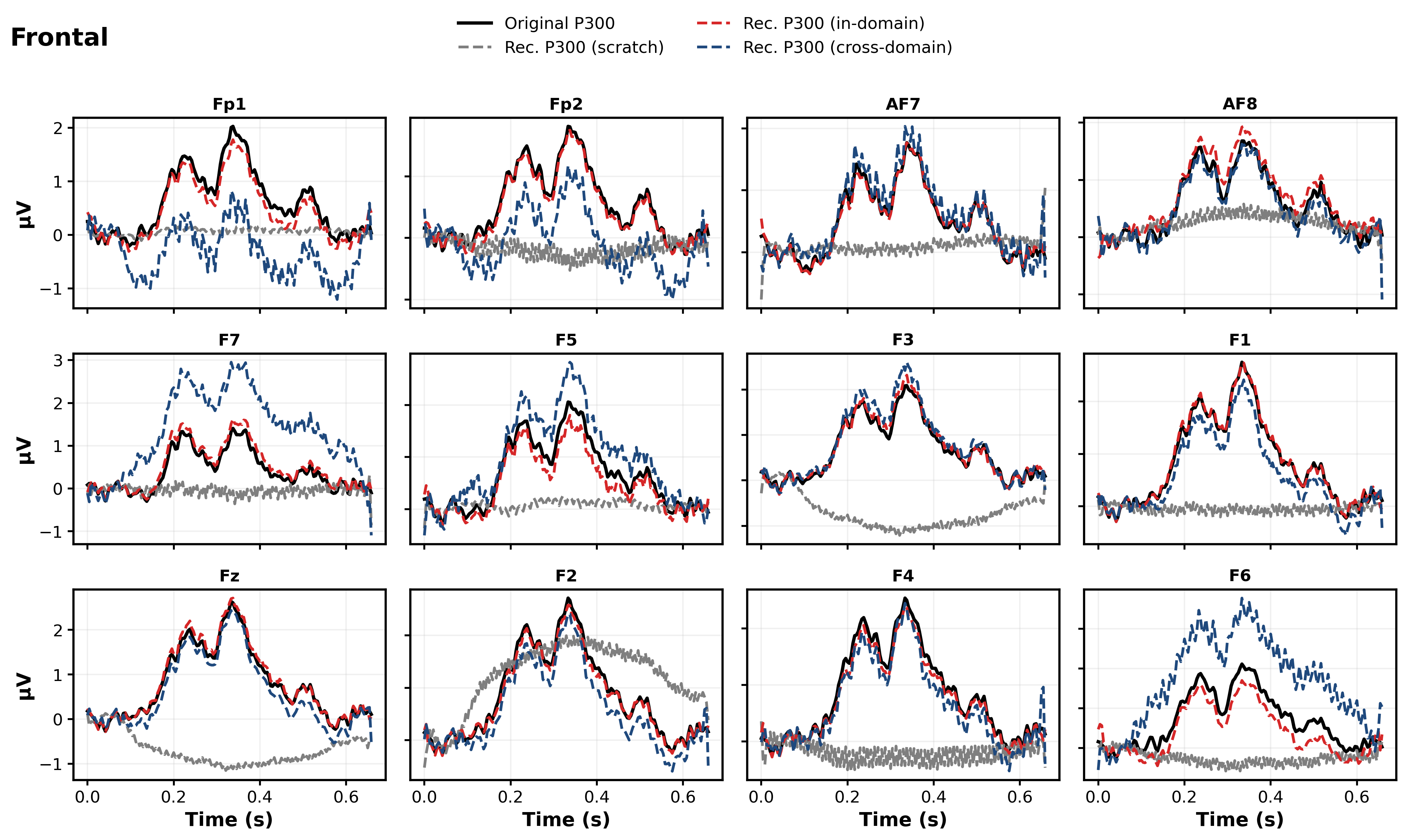}
  \includegraphics[width=0.9\linewidth]{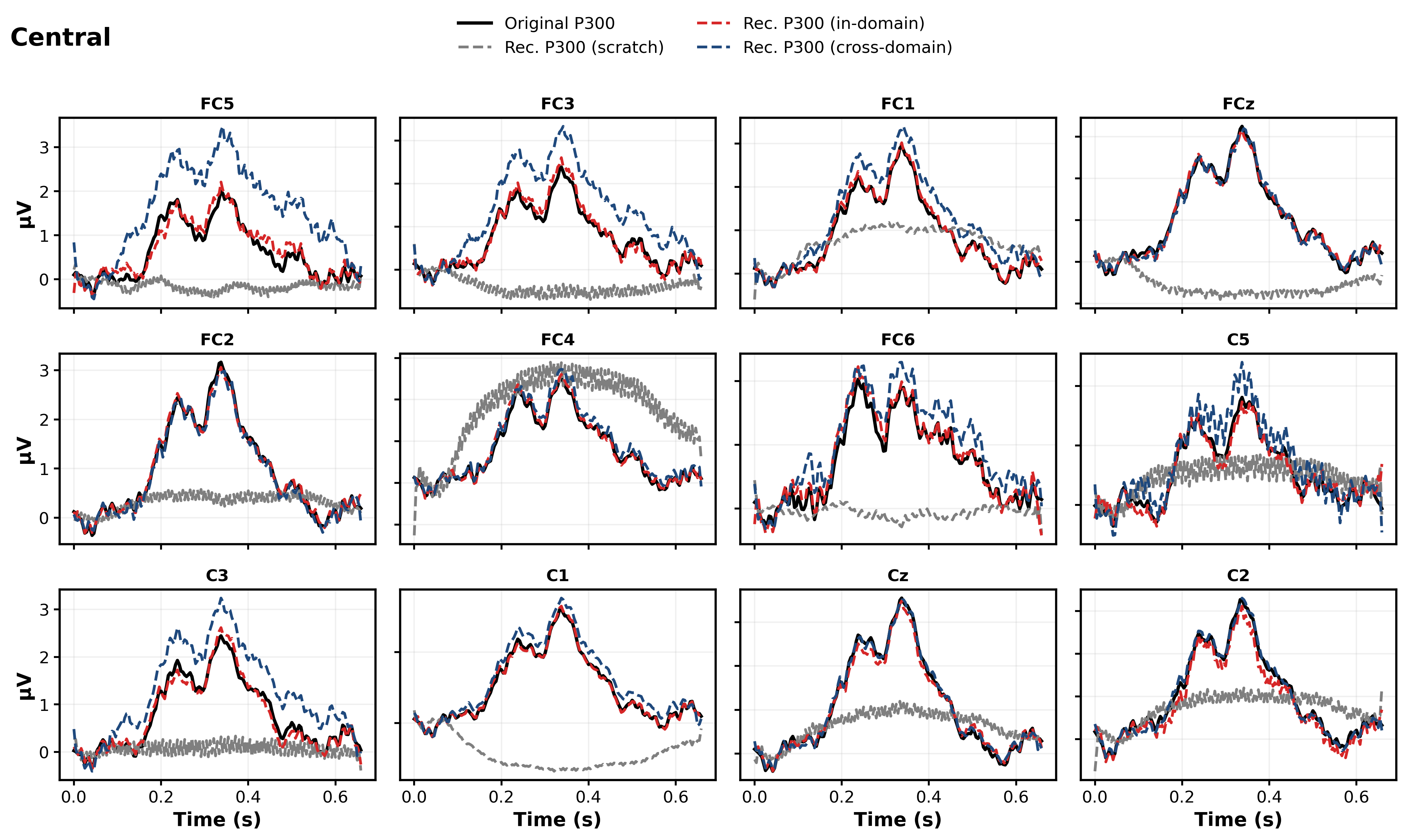}
  \caption{Region-wise reconstruction of P300 ERPs for (a) Frontal and (b) Central regions.}
  \label{fig:erp_regions1}
\end{figure}

\begin{figure}
  \centering
  \includegraphics[width=0.9\linewidth]{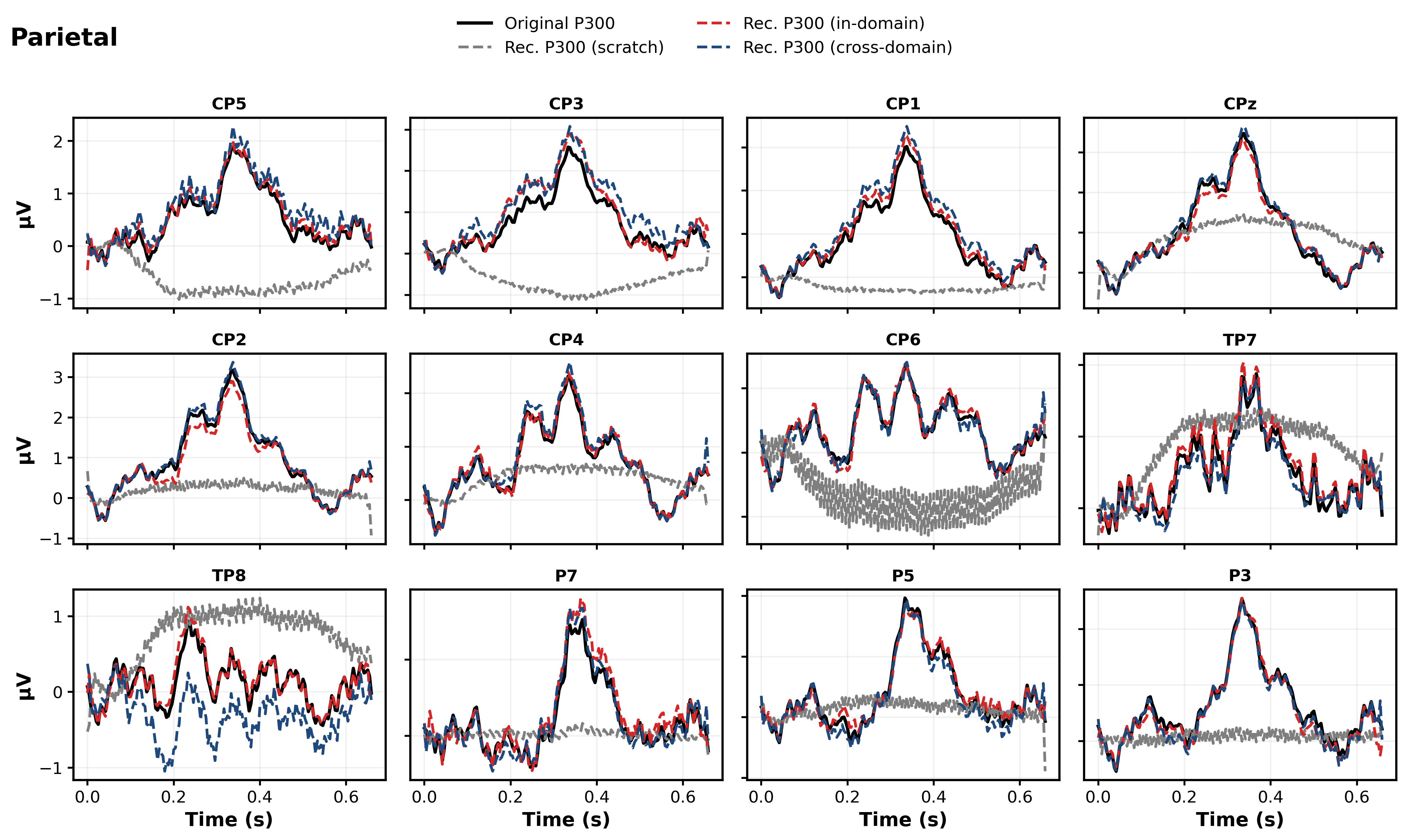}
  \includegraphics[width=0.9\linewidth]{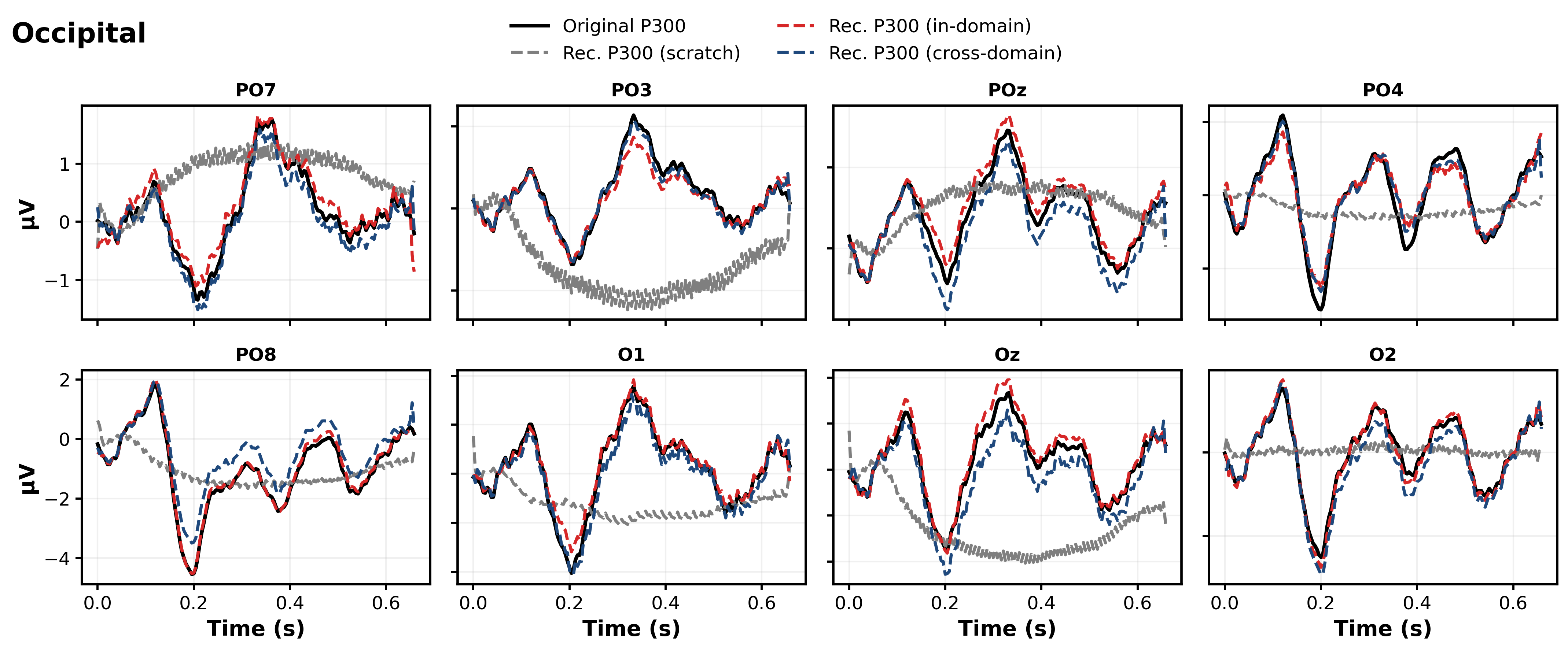}
  \includegraphics[width=0.9\linewidth]{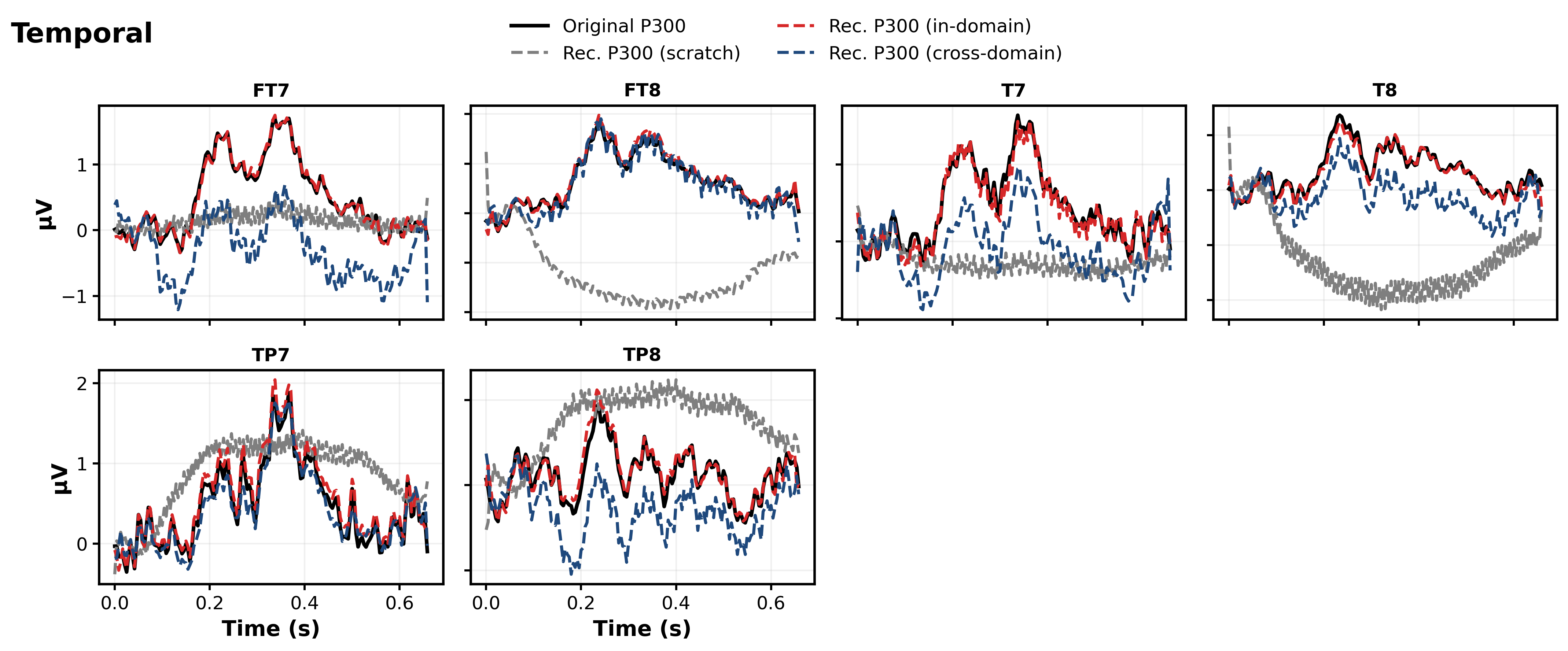}
  \caption{Region-wise reconstruction of P300 ERPs for (c) Parietal, (d) Occipital, and (e) Temporal regions.}
  \label{fig:erp_regions2}
\end{figure}

To complement Figure~\ref{fig:erp_regions1} and ~\ref{fig:erp_regions2}, which provides an overview of reconstruction performance across all 64 EEG channels, we further present detailed visualizations of event-related potentials (ERPs) grouped by brain region.  
For each region (Frontal, Central, Parietal, Occipital, and Temporal), we display up to 12 representative channels in a grid layout.  
Each subplot shows the original averaged P300 response (black) compared with reconstructions from three models: trained from scratch (gray), in-domain pretraining (red), and cross-domain pretraining (blue).  

These region-wise plots highlight differences in reconstruction fidelity across scalp locations.  
In particular, in-domain pretraining achieves reconstructions closely aligned with ground truth across most regions, while cross-domain pretraining captures the overall waveform morphology but may underestimate amplitudes in some frontal and occipital sites.  
Scratch models generally fail to recover meaningful P300 waveforms. Together, these results demonstrate the robustness and transferability of SpellerSSL across spatially distributed brain regions.

\begin{figure*}
  \centering
  \begin{subfigure}{0.48\linewidth}
    \centering
    \includegraphics[width=\linewidth]{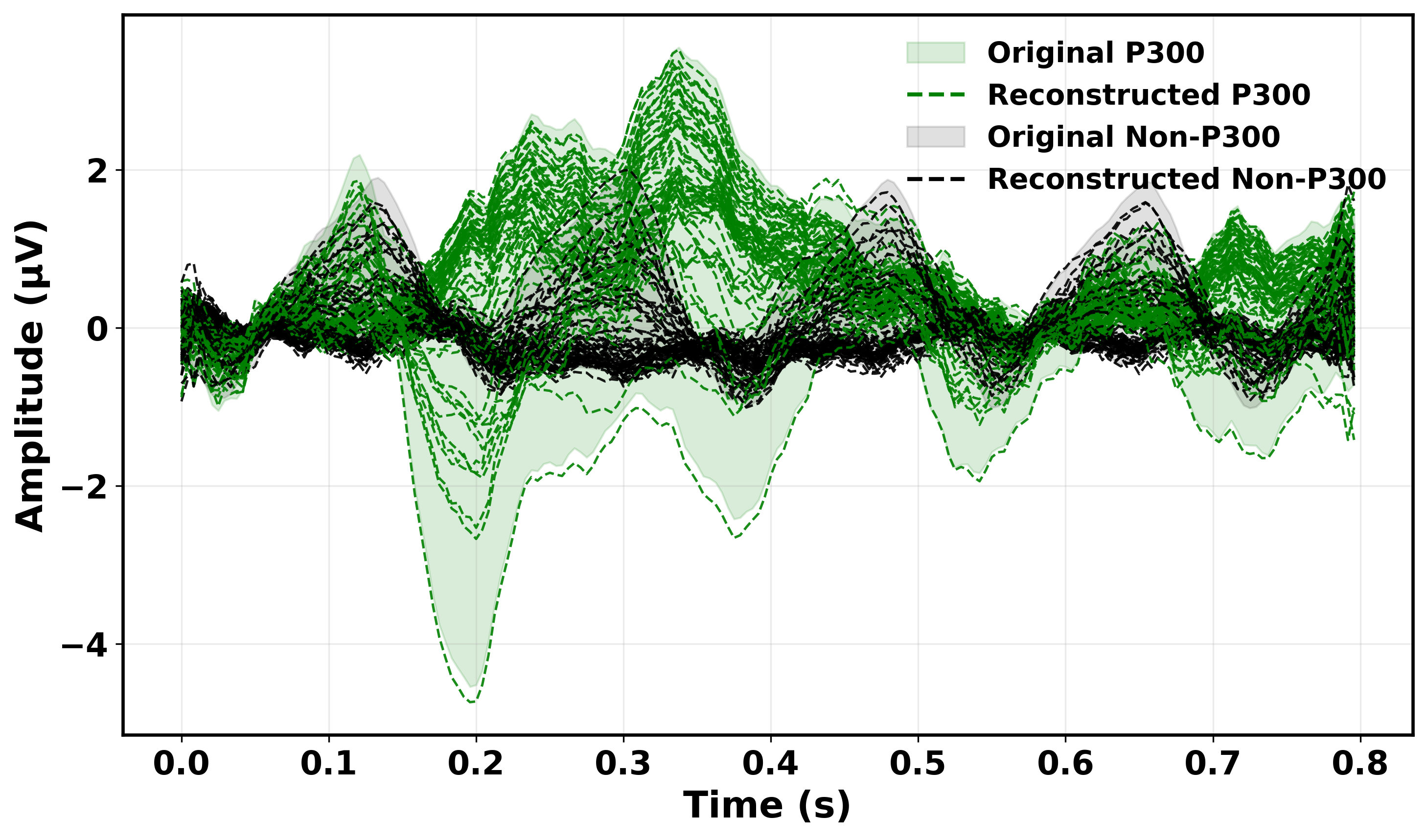}
    \caption{In-domain ERP across all channels.}
    \label{fig:BG1}
  \end{subfigure}
  \begin{subfigure}{0.48\linewidth}
    \centering
    \includegraphics[width=\linewidth]{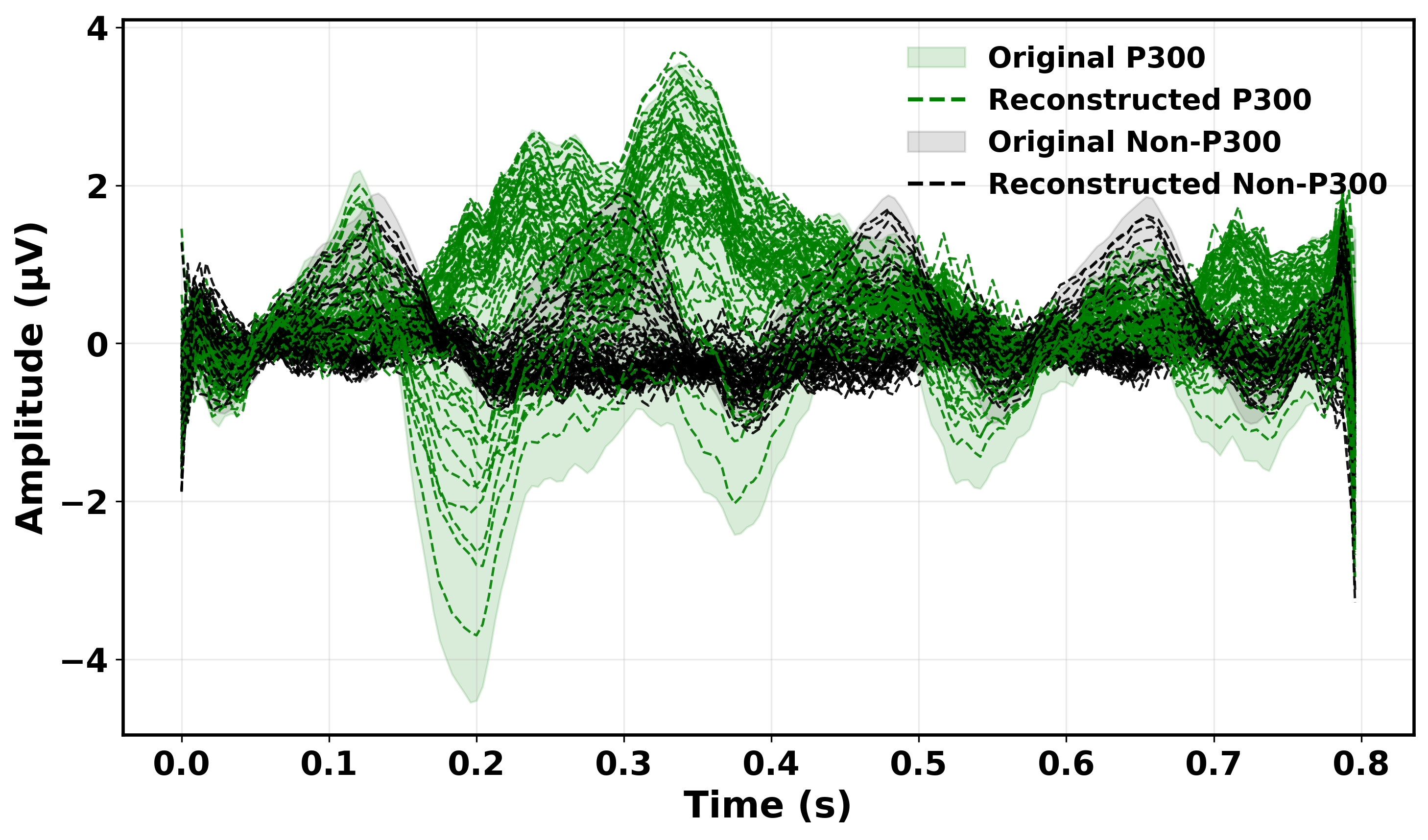}
    \caption{Cross-domain ERP across all channels.}
    \label{fig:BG2}
  \end{subfigure}
  \caption{Overall reconstruction visualizations of P300 and Non-P300 ERPs across all 64 channels under cross-domain and in-domain pretraining. Original responses (solid lines) are shown with reconstructions (dashed lines), with shaded regions indicating variability across channels.}
  \label{fig:overall_recon}
\end{figure*}

In addition to the region-wise ERP comparisons, we also provide unified reconstructions across all channels to illustrate overall signal morphology. 
Figure~\ref{fig:overall_recon} shows grand-average visualizations of P300 and Non-P300 responses under cross-domain and in-domain pretraining. 
Original signals are plotted alongside their reconstructions, with shaded regions denoting variability across channels. 
These plots complement the channel- and region-specific figures.

\section{Additional Baseline Results}\label{app:baseline_crr}
\begin{table}
\centering
\caption{Comparison of CRRs (\%) between SpellerSSL and state-of-the-art models.}\label{tab:CRR}
\resizebox{0.9\linewidth}{!}{
\begin{tabular}{lc|cccccccccccccc c}
\hline
\multicolumn{1}{l|}{Model}          & Ref.    & 1           & 2           & 3           & 4           & 5           & 6           & 7           & 8           & 9           & 10          & 11          & 12          & 13          & 14          & 15          \\ \hline
\multicolumn{1}{l|}{ST-CapsNet}     &   \cite{wang2023st}      & 41          & 61          & 66          & \textbf{78} & \textbf{85} & 86          & 92          & 90          & 91          & 95          & 96          & 96          & 95          & 97          & 96          \\
\multicolumn{1}{l|}{ERP-CapsNet}    &    \cite{ma2021capsule}     & 45          & 60          & 66          & 73          & 81          & 85          & 87          & 90          & 90          & 94          & 94          & 94          & 94          & 95          & 96          \\
\multicolumn{1}{l|}{CNN}          &    \cite{cecotti2010convolutional}     & 35          & 52          & 59          & 68          & 79          & 81          & 82          & 89          & 92          & 91          & 91          & 90          & 91          & 92          & 92          \\
\multicolumn{1}{l|}{MCNN}         &   \cite{cecotti2010convolutional}      & 39          & 55          & 62          & 64          & 77          & 79          & 86          & 92          & 91          & 92          & 95          & 95          & 95          & 94          & 94          \\
\multicolumn{1}{l|}{S-LDA}          &   \cite{blankertz2011single, ma2021capsule}      & 39          & 54          & 64          & 67          & 75          & 78          & 85          & 88          & 87          & 92          & 92          & 96          & 93          & 94          & 96          \\
\multicolumn{1}{l|}{SWFP}           &    \cite{gerson2006cortically, wang2023st}     & 40          & 60          & 69          & 71          & 80          & 82          & 86          & 89          & 89          & 92          & 92          & 93          & 92          & 94          & 94          \\
\multicolumn{1}{l|}{P3T}            &    \cite{hong2024p3t}     & 42          & 62          & \textbf{74} & 72          & 81          & 85          & 90          & 91          & 92          & 92          & 94          & 94          & 94          & 94          & 96          \\ \hline
\multicolumn{2}{l|}{Cross-domain}     & \textbf{44} & \textbf{65} & \textbf{74} & 75          & 82          & 87          & 93          & \textbf{92} & \textbf{95} & 95          & 96          & 96          & 96          & 97          & \textbf{97} \\
\multicolumn{2}{l|}{In-domain}        & 43          & \textbf{65} & 69          & 74          & 84          & \textbf{89} & \textbf{94} & 91          & 93          & \textbf{96} & \textbf{98} & \textbf{99} & \textbf{98} & \textbf{99} & \textbf{97} \\
\multicolumn{2}{l|}{In-domain (60\%)} & 37          & 59          & 72          & 77          & 82          & 87          & 91          & 92          & 93          & \textbf{96} & 96          & 97          & 96          & 97          & \textbf{98} \\ \hline
\end{tabular}
}
\end{table}

Table~\ref{tab:CRR} reports a detailed comparison of character recognition rates (CRR, \%) between SpellerSSL ($G=2$) and representative state-of-the-art P300 speller models across repetitions 1–15. 
These results complement the ITR analysis in the main text (Table~\ref{tab:itr}), providing an additional insight into recognition accuracy over increasing repetitions. CRRs above 90\% are underlined, and the best values at each repetition are highlighted in bold. 

Compared with classical approaches (e.g., S-LDA, SWFP) and recent deep models (e.g., ERP-CapsNet, ST-CapsNet, P3T), SpellerSSL achieves superior or comparable performance across most repetitions. Notably, early recognition (repetitions 1–3) is improved by SSL pretraining, with cross-domain and in-domain checkpoints showing faster accuracy gains than baselines. 
At later repetitions, in-domain SpellerSSL maintains the highest CRRs (up to 99\%), confirming the benefit of self-supervised pretraining not only in terms of ITR efficiency but also in raw recognition accuracy. 

These results further support the conclusion that SSL pretraining enhances both speed and accuracy, and that SpellerSSL can match or surpass existing state-of-the-art methods even under reduced calibration (e.g., in-domain 60\%).

\end{document}